\documentclass[oneside,10pt,a4paper]{article}      
\usepackage{amsmath}
\usepackage{amssymb}
\usepackage{graphicx}

\usepackage{lineno}

\usepackage[margin=16mm]{geometry}
\title{\bf Transient nutations of an electron spin 1/2 with dipolar coupling to neighbouring nuclear spins}
\author{Alain Deville, Anne-Marie Dar\'e, Julien Azema \\
\vspace{1mm}\\
Aix-Marseille Universit\'e, CNRS, IM2NP UMR 7334, 13397, Marseille, France} 
\date{}      

\begin{document}             

\maketitle                   

\begin{abstract}
\normalsize A model cluster, with a resonant electron spin surrounded with identical nuclear spins occupying the sites of a cubic lattice, and with dipolar coupling between the spins, is used for a theoretical investigation of the transient nutations induced by a strong oscillating field but disturbed by these local dipolar fields. An effective Hamiltonian valid at the time scale of the nutations is established. An analytical expression for the Dipolar Coherence time $ \mathit{ t_{DC}} $ is derived. The dipolar coupling is found not to affect t$_{DC}$ at exact resonance ($\Delta =0)$, a result reminiscent of a behaviour found in transient NMR in liquids, in an inhomogeneous static field. The expression for t$_{DC}$ also presents a qualitative similarity with a time introduced in that liquid-state NMR context. Introducing a collection of independent clusters, a bridge is built between $\mathit{ t_{DC}} $ and the half-width $\delta $ of the usual steady state unsaturated ESR line. Numerical values for $\mathit{(t_{DC})_{\Delta = \delta}}$ are given, for undiluted and dilute samples. It is shown that the TN do not obey the Bloch equations. Precautions to be taken when trying a qualitative comparison between the present results and existing experimental data, and some thermodynamic aspects, are discussed. 
\end{abstract}
\section{Introduction}\label{Introduction}
Transient Nutations (TN) may be induced by a magnetic field in Magnetic Resonance spectroscopy (NMR \cite{Torrey1949} , ESR \cite{Atkins1974}), or by an electric field in optical  spectroscopy  \cite{Hocker1968}. In ESR, if the electron spins are at thermal equilibrium in a static resonant field when a pulse of oscillating field starts on, then, in addition to the appearance of the usual absorption, a finer oscillatory phenomenon at the Rabi frequency \cite{Rabi1937,Abragam1961} immediately develops.  The Rabi frequency is quite smaller than that of the oscillating field. This phenomenon is clearly visualized within the effective field concept (cf. Section \ref{SpinsAEtB-Modele}). In \cite{Atkins1974}, as in \cite{Torrey1949}, the host of the magnetic moments was a liquid, where the steady state resonance lines are narrowed by the molecular motions, which nearly average out the dipolar coupling between spins \cite{Abragam1961}. In solids, motional narrowing is absent but, at high electron spin concentration, exchange coupling may, sometimes drastically, reduce this dipolar field and its effect. Exchange narrowing was present in the paramagnetic salts of TCNQ studied with the transient nutations method by Berthet \textit{et al} \cite{Berthet1976}. The observation of these forced nutations in ESR is generally difficult, as e.g. it must be made during the pulse of intense oscillating field, which may destroy or at least transiently saturate the reception part of the ESR spectrometer. A 1993 paper \cite{Boscaino1993} identifies these difficulties in detail. Most often, such an observation is therefore made only indirectly, through Free Induction Decay (FID) and/or spin echoes, an approach which has allowed the creation of specific spectroscopic techniques \cite{Astashkin1990, Fedoruk2002, Schweiger2001}. A 2015 ArXiv document \cite{Asadullina2015} identifies only two papers \cite{Boscaino1993, Agnello1999}  with direct observation and measurement of these TN, made possible through two-quantum excitation. A third paper is \cite{Berthet1976}.\\
The present paper focuses upon a paramagnetic impurity surrounded by nuclear spins. In this situation, the NMR of the neighbouring nuclear spins, as opposed to the ESR of the central electron spin, was examined by Bloembergen as early as 1949 \cite{Bloembergen1949}, experimentally between 1 and 300 K and theoretically. He showed that the Spin-Lattice (SL) relaxation of the nuclear spins was controlled by the electron spins, even at a $10^{-6}$ concentration of electron spins. The role of the diffusion of the nuclear magnetization in that SL process was later identified \cite{DeGennes1958, Blumberg1960, Abragam1961} (see Sections \ref{startingHamiltonian} and \ref{UseandAbuse}). We are presently interested in the TN of the electron spin, motivated by four facts: 1) the foundations of Quantum Theory are still under debate and stimulate experimental investigations about the measurement process \cite{Wheeler1983, Schlosshauer2004} and the decoherence phenomenon \cite{Schlosshauer2004, Buchleitner2009}. 2) In the context of Quantum Information Processing (QIP), efforts in Quantum Computing (QC) are specifically promoting the use of nuclear or electron spins in the physical realizations of qubits and quantum logic gates \cite{Nielsen2000}. 3) Technological progress promotes the development of tools \cite{Franck2015}  which could e.g. help in a direct observation of the TN of electron spins. 4) the existing results upon the TN of electron spins in insulators generally come from rather complex situations. A theoretical treatment of the present simpler case could be of help in this context.\\
 The following conditions are assumed in this paper: frequency of conventional ESR spectrometers (9-35 GHz), conventional temperature range (1-300 K), electron spin associated with a localized wave-function, as found with paramagnetic ions in diamagnetic insulators \cite{Abragam1971}. A donor centre in GaAs at low temperature (effective Bohr radius  $a_0^{*} \simeq 100$ \AA, cf. Eq. (9) of \cite{Paget1977}) would be an opposite situation, the antisite defect in amorphous GaAs \cite{Deville1989} an intermediate one. The nucleus of a possible central paramagnetic ion is assumed spinless.\\
It is an experimental fact that, in the room-temperature range and possibly at lower temperatures, some paramagnetic defects cannot be detected with ESR, because they are so strongly coupled to the lattice that the ESR line is wider than the ESR frequency. We presently consider an opposite situation, when the spin-lattice relaxation time is far longer than the time associated with the dipolar coupling between an electron moment and the neighbouring nuclear moments. We are interested in a transient regime, the amplitude of the pulse of microwave field being far greater than that of the dipolar field on the electron spin.  In section \ref{SpinsAEtB-Modele}, the validity of a spin temperature or a spin bath assumption is discussed, an approximate effective hamiltonian for the spin moments valid at the time scale ot the   TN is established, through two successive familiar unitary transformations. and it is explained why general results from linear response theory, from the study of relaxation phenomena, or from the field of steady state saturation cannot be taken for granted during this pulse. For these reasons, the behaviour of the electron spin during the pulse is thoroughly examined in the following sections. Two recent papers, from Dobrovitski \textit{et al} \cite{Dobrovitski2009} and from Baibekov \cite{Baibekov2011}, examined theoretically the  TN of a central electron spin with dipolar coupling to neighbouring spins. In \cite{Dobrovitski2009} the  neighbours were non resonant electron or  nuclear spins, and the amplitude of the transient field was assumed far greater than the local dipolar field. The treatment in \cite{Dobrovitski2009} was largely phenomenological, as stressed in \cite{Baibekov2011}. Here, we will use first principles for finding first an approximate expression of the central spin reduced density operator, and then a time characterizing the TN. We will then relate this time to the ESR unsaturated linewidth (linear regime) calculated with the well-established Van Vleck method of moments \cite{VanVleck1948}. In fact, the statistical approach used in \cite{Dobrovitski2009} dates back to Klauder and Anderson \cite{Klauder1962}, who introduced it in the context of the linewidth in MR, when the method of moments is too cumbersome. Presently, with neighbouring nuclear spins, the method of moments happens to be tractable, and should therefore be preferred, as it is more accurate. Baibekov used first principles in \cite{Baibekov2011}, but he  considered a different and more difficult situation, with a single spin species.\\
 From now on, a collection of electron spins of species A, each surrounded by nuclear spins of species B in an otherwise diamagnetic solid, will be considered. The size of a $\{$Central A - neighbouring B spins$\}$ cluster is high enough to present negligible surface effects and negligible coupling with adjacent clusters. Exchange coupling is then absent. When discussing the TN, or the ESR unsaturated linewidth, a single cluster may then be considered. In both cases, if N electron spins are present, the total intensity is N times the intensity from a single cluster. For the sake of simplicity, all spins are assumed to be 1/2. If A, the central spin, is associated with a paramagnetic ion, this assumption is moreover adapted to the QC context, when qubits are electron spins. In Section \ref{TransientBehavior}, the expression of the reduced density operator for the central electron spin is first established, which then allows to find the time-dependence of the mean value of its components. These expressions are used for numerical calculations when the nuclear neighbours are distributed over the sites of a cubic lattice. The results are discussed in Section \ref{SectionDiscussion}. 
\section{Towards an effective Hamiltonian for the TN of the central spin}\label{SpinsAEtB-Modele}
In this section, first the description of the system is completed. As most quantum problems, the transient forced regime discussed in this document has no exact analytical solution, and approximations are therefore needed. It is explained why, at the time scale of the TN, the dipolar coupling between the nuclear moments can be neglected, and a spin temperature for these nuclear moments would be unrealistic. It is recalled why treating the nuclear spins as a bath for the electron spins would be even more demanding.
\subsection{The starting Hamiltonian at the time scale of the TN}\label{startingHamiltonian} 
An electron spin A (magnetic moment $\boldsymbol{\mu}$ =$-g_{e} \mu_{B} \textbf{\textit{s}}$, $ \mu_B $: Bohr magneton), is surrounded by $ N $ nuclear spins of species B (magnetic moment for spin j: $\boldsymbol{\mu}_j$ = 
$-g_{n}\mu_{N}\boldsymbol{\mathit{I_j}}$, $\mu_N $: nuclear magneton), all being spins $ 1/2 $. These spins are submitted to a static magnetic field $B_0 \boldsymbol{\mathit{k}}$ (amplitude $B_0 $, direction $ z $ chosen as the quantization axis). In the present situation, as defined in Section \ref{Introduction}, any hyperfine contact term in the spin Hamiltonian for these spins is absent. The spin Hamiltonian consists of the Zeeman term only, written as $\mathcal{H_Z}=\mathcal{H}_{Ze}+\mathcal{H}_{Zn}$,  with $\mathcal{H}_{Ze}=\hbar\omega_{0}s_z $ and $ \mathcal{H}_{Zn}=-\hbar\omega_{0n}I_z,$ with $I_z=\Sigma_jI_{jz} $ ($\omega_{0}=g_e \mu_B B_0/\hbar$, $\omega_{0n}=g_n \mu_N B_0/\hbar$ : resonance frequencies for A and the B spins respectively). From $ t=0 $ onwards, the spins are moreover submitted to an intense microwave magnetic field $ 2B_1\cos\omega t \boldsymbol{i}$ with linear polarization (easier to realize experimentally than a circularly polarized one), along the $ x $ axis, with amplitude $ 2B_1, $ and frequency $ \omega $ close to $ \omega_{0} $ (transient ESR experiment). The energy of coupling of A with the resonating rotating component of the oscillating field is denoted as $ \hbar \omega_1 = g_e \mu_B B_1$. At X-band, it is possible to get $2B_1\simeq 1$ mT (cf.  Section \ref{SectionDiscussion}). The Hamiltonian for the dipolar coupling between two point magnetic moments $\boldsymbol{\mu_i}$ and $\boldsymbol{\mu_j}$, at a distance $ r_{ij} $, is
\begin{equation}
(\mathcal{H}_\mathit{d})_{ij}=\frac{\mu_0}{4\pi}(\frac{\boldsymbol{\mu_i}\boldsymbol{\mu_j}}{r_{ij}^3}-3\frac{(\boldsymbol{\mu_i}\boldsymbol{r_{ij}})(\boldsymbol{\mu_j}\boldsymbol{r_{ij}})}{r_{ij}^5}).
\end{equation}
A nuclear B spin in a state $m_{I}$ creates a local field with a maximum  amplitude  $B_{n}\simeq (\mu_{0}/{4 \pi})(g_{n}\mu_N/r^3)\mid m_{I} \mid$ on any spin at a distance $r$. 
For proton neighbours ($g_{n}=5.586$) and $r=3$ \AA, $B_{n}=0.05$ mT. Just before $ t=0 $, the spins are at thermal equilibrium, or have been prepared in some state. Our aim is to characterize the time for the decay of the nutations induced by the microwave pulse. The SL coupling for the central electron spin (corresponding characteristic time $T_{1e}$) is supposed to be far weaker than the dipolar coupling between A and its B nearest neighbours (n.n) (choice of the paramagnetic centre and/or temperature), and we therefore ignore the SL coupling of spin A during these nutations. 
In Section \ref{Introduction} it was said that the dipolar coupling between nuclear spins is responsible for the SL nuclear relaxation (time $T_{1n}$), through the central electron spin A (diffusion of nuclear magnetization, and SL relaxation of A), and therefore $T_{1n}^{-1}$  $<$ $T_{1e}^{-1}$. We must then also ignore this indirect SL of the B nuclear spins. We thus presently ignore the SL phenomena (formally, infinite $T_{1e}$ and  $T_{1n}$)). The decay of these nutations is then due to the dipolar coupling between A and the neighbouring nuclear spins. An examination of \cite{Blumberg1960} moreover shows that, even with samples and time scales allowing nuclear spin diffusion, the immediate neighbours of the central electron spin do not take part in the process: existence of a diffusion barrier, modeled with a diffusion sphere around the electron moment. The present physical situation is then the following: the time for the duration of the transient nutations is imposed by the magnetic dipolar energy between an electron moment and a nearby far weaker (at least a factor of $10^{3}$) nuclear moment. A determination of that time is made in the following sections. The diffusion phenomenon is far more slower, as it implies a collection of the, weak, nuclear moments. Since, in the presence of paramagnetic impurities, the diffusion phenomenon within the nuclear spins may operate in the thermal relaxation of the nuclear spins through electron spins, one may have a more quantitative idea of that diffusion time, thanks to Abragam and Goldman (page 135 of their canonical book \cite{Abragam1982}): in insulating solids, and D being the diffusion constant : \textit{\textquotedblleft The magnitude of D for a diffusion induced by dipolar interaction is exceedingly small, of the order of $10^{-12} cm^{2}s^{-1}$ and the propagation of magnetization exceedingly slow. Over a distance $r$ it requires a time $\tau \simeq r^{2}/D$, of the order of several hours for one micron\textquotedblright.} It is therefore legitimate to neglect the dipolar coupling between the nuclear spins, compared with the one between the electron spin and the nuclear spins, and to write the following Hamiltonian:
\begin{equation}\label{PourHamiltonienDeDépart}
\mathcal{H_Z} + 2\hbar\omega_{1}(s_x -\frac{\omega_{0n}}{\omega_{0}}I_x)\cos\omega t +\mathcal{H}_{den},
\end{equation}
where $\mathcal{H}_{den}$ =$\sum_j \mathcal{H}_{dj}$ is the dipolar Hamiltonian between A and the B spins. In the present situation, the ESR line from a collection of independent clusters, with a width due to dipolar inhomogeneous broadening by \textit{nuclear} spins, is narrow, and only one of the two rotating components of the oscillating field is effective and will be kept. Moreover, the quantity $(-\omega_{0n}/ \omega_0)$ $I_x)\cos\omega t$ corresponds to a weak $(\omega_{0n} << \omega_0)$ and non-resonant nuclear term, and will not be kept either. The analysis of the TN thus starts with the following approximate Hamiltonian $\mathcal{H}$:
\begin{equation}\label{HamiltonienDeDépart}
\mathcal{H} =\mathcal{H_Z} + \hbar\omega_{1}(s_x \cos\omega t + s_y \sin\omega t)+\mathcal{H}_{den}.
\end{equation}
We follow \cite{Abragam1961}, and write $\mathcal{H}_{dj}$ as:
\begin{equation}\label{DefinitionAtoD} 
\mathcal{H}_{dj} =f(r_j)(A_j+B_j+C_j+C_j^\dagger+E_j+E_j^\dagger),
\end{equation}
with 
\begin{eqnarray}
f(r_j)&=&\frac{\mu_0}{4\pi}\frac{g_eg_n\mu_B\mu_N}{r_{j}^3}, \\
A_j &=&(1-3\cos ^2\xi_j)s_zI_{jz}, \quad B_j =-\frac{1}{4}(1-3\cos ^2\xi_j)(s^{+}I_j^{-}+s^{-}I_j^{+}),  \label{DipolairesA} \\
C_j &=&-\frac{3\sin \xi _{j}\cos \xi _{j}e^{-i\varphi _{j}}}{2}(s_zI_j^+ + s^+ I_{jz}), \quad E_j =-\frac{3\sin ^2\xi_je^{-2i\varphi _j}}{4}s^+ I_j^+ , \\ \nonumber
\end{eqnarray}
where $\xi _j$ and $\varphi _j$ are the polar angles of $\boldsymbol{r_j}$. Defining $r_{\min }$, the shortest distance between A and a B spin, then from the assumptions made:
\begin{equation}\label{ConditionDobrovitski}
f(r_{min })<<\hbar\omega _1<<\hbar\omega _{0},
\end{equation}
We will call (\ref{ConditionDobrovitski}) the weak dipolar coupling or Redfield condition, as it was used in \cite{Redfield1955}. In his canonical paper \cite{VanVleck1948} (cf. also Ch.IV of \cite{Abragam1961}), Van Vleck did consider dipolar-broadened lines, in solids, but he focused upon an usaturated line (linear regime) in a Continuous-Wave (CW) experiment. There, the energy of coupling with the rotating field was far weaker than the dipolar energy. We will call the inequalities $\hbar\omega _{1}$ $<<$ $f(r_{min})$ $<<\hbar\omega _{0e}$ the intermediate dipolar coupling or Van Vleck condition. Consequently, it is presently impossible just to refer to the results from \cite{VanVleck1948} or Ch.IV of \cite{Abragam1961}. In the study of relaxation phenomena (cf. Ch. VII and IX of \cite{Abragam1961}), the energy of coupling to the rotating field is again far weaker than the dipolar energy, and the corresponding results are also presently useless.
\subsection{Use and abuse of thermodynamic assumptions}\label{UseandAbuse} 
It could happen that some readers unfamiliar with ESR now introduce a spin temperature for the nuclear moments, or even treat them as a bath for the electron spins. But information already given in Sections \ref{Introduction} and \ref{startingHamiltonian}, and the time scale of the TN, suggest that these assumptions are unrealistic, as detailed below.\\
The spin temperature concept, present in early papers from Casimir, Du Pr{\'{e}}, Van Vleck (see \cite{VanVleck1940} and references therein) was developed by Abragam and Proctor \cite{Abragam1958}. Both Abragam and Goldman (see e.g. \cite{Goldman1970, Abragam1982}) stressed that speaking of a spin temperature is an assumption, which has to be validated experimentally in each situation. It is associated with an internal equilibrium within a spin system, not with its equilibrium with a bath (or thermostat). Redfield could even extend it to an internal equilibrium within the so-called rotating frame, in a stationary regime in saturation conditions, which allowed him to successfully explain the apparently anomalous behaviour of dispersion, in NMR, in metals \cite{Redfield1955}, and this extension was soon validated in other nuclear systems \cite{Goldburg1961, Solomon1962}. Presently, such a spin-temperature assumption for the nuclear spins during the TN would be unrealistic, for at least two reasons: 1) the electrons are resonating, not the nuclei. 2) as explained in Section \ref{startingHamiltonian}, diffusion affects the nuclear spins outside the diffusion barrier only, and this process needs a time far longer than the duration of the TN. On the contrary, the nuclear spins near an electron spin are directly affected by the dipolar field from the electron moment, a phenomenon to be described explicitly. Then, during the TN,  it  would be totally unjustified to assume the existence of a spin temperature for the nuclear spins inside the diffusion barrier centered on an electron spin, and/or for those outside this barrier. This argument 2) is another reason why, in this study of the TN, the dipolar coupling between nuclear spins should be neglected compared with the one between the electron and the nuclear spins.\\
It would even be more unrealistic to presently assume that, at the time scale of the TN, the nuclear spins are a bath (or thermostat) for the paramagnetic impurities. Such an assumption is far stronger than that of a spin temperature, since the following conditions should all be fulfilled: 1) energy should be transferred from the electron to the nuclear spins during the nutation step. But one may guess and it will shortly be confirmed numerically that the dipolar coupling links a given electron spin to its neigbouring nuclear spins only. Moreover, in a first order perturbation approach, the $s^{\pm}I^{\mp}_{j}$ dipolar terms do not induce any energy transfer between an A spin and its B nuclear neighbours during the nutation step, and going into higher perturbation theory would not significantly change the situation, because of the important mismatch between electronic and nuclear Zeeman energies. 2) the nuclear spins should be coupled to the lattice (presently the phonons), but this coupling is quite weak (their $T_1$ may typically be from seconds to hours \cite{Abragam1961}). 3) the nuclear spins should have to get an internal equilibrium in a time far shorter than the duration of the TN. But it has just been explained that, on the contrary, they cannot get a spin temperature during the TN. 4) the heat capacity of the nuclear spin system should be far greater than that of the electron spins, which is not presently the case, as only the neighbouring nuclear spins of a given paramagnetic impurity is significantly coupled to this electron spin, and a nuclear moment is far smaller than an electron moment. As a consequence, if the B neighbours are nuclear spins in \cite{Dobrovitski2009}, the assumption that they act as a thermal bath should not be kept.\\
The short duration of the TN was one of the reasons just given for presently excluding any spin bath or spin temperature assumption for the nuclear spins during these TN. It should be added that the pulse of intense oscillating field, at or near the resonance field of the electron spin, which creates the nutations, creates non-diagonal elements in the density matrix, which makes any temperature approximation inappropriate during these nutations \cite{Abragam1961}. In the context considered by Redfield when he postulated the existence of a spin temperature in the rotating frame (cf. \cite{Redfield1955} and Ch.XII of \cite{Abragam1961}), the weak dipolar coupling condition is satisfied,  but the results cannot be presently used, because they refer to a stationary regime (the spin temperature does not establish spontaneously). And moreover, even in a stationary regime, when Redfield mentioned the ESR saturation of F centres surrounded by nuclear spins, in alkali halides \cite{Redfield1955}, he stressed that although \textquotedblleft the different F-centers in an alkali halide are magnetically coupled in theory via nuclear spin diffusion and direct interaction with each other, in fact such coupling is negligible over a length of time comparable to $T_{1}$ (electronic)", which forbids any assumption of a spin temperature in the (ESR) rotating frame. For all these reasons, it is also presently impossible just to start with results from \cite{Redfield1955}.
\subsection{The effective Hamiltonian}
When the Van Vleck condition is fulfilled, then from time-independent perturbation theory (cf. Chap. IV of \cite{Abragam1961}), one has to keep only the terms of the dipolar Hamiltonian which commute with the Zeeman Hamiltonian, called its secular part, i.e.  only the $\mathit{A}$ term in Eq. (\ref{DefinitionAtoD}) when, as presently, there are two spin species and only one is resonating. As the Van Vleck condition is presently not fulfilled, this truncation through time-independent theory is not possible, and one must start from the full Hamiltonian $\mathcal{H}$, Eq. (\ref{HamiltonienDeDépart}). In order to eliminate uninteresting parts of  the dynamics, we introduce  $U_1$ = $e^{i\omega t (s_z+I_z)}$ and the unitary transformation $\mid \Psi_{1}(t)>$ = $U_1$ $\mid \Psi(t)>$ \cite{Redfield1955}. $\mid \Psi_{1}(t)>$  obeys the Schr\"{o}dinger equation $i\hbar(d/dt)\mid \Psi_{1}(t)>$=$\mathcal{H}_1$ $\mid\Psi_{1}(t)>$, where:
\begin{eqnarray}
\mathcal{H}_1&=&i\hbar\frac{dU_1}{dt}  {U_{1}^+}+U_1\mathcal{H}U_{1}^{+},\label{hamiltonienH1} \\ 
&=&\hbar \Delta s_z+\hbar \omega_1 s_{x}+\hbar \Delta_n I_z+U_1 \mathcal{H}_{den} U_{1}^{+},
\end{eqnarray}
with $\Delta=\omega_0-\omega$ and $\Delta_n=-\omega_{0n}-\omega$ $\simeq$ $-\omega$. The quantity $\hbar \Delta s_z+\hbar \omega_1 s_{x}$ is the Zeeman energy of the central spin in an effective static field $\boldsymbol{B_{eff}}$ obeying $g_e\mu_B\boldsymbol{B_{eff}}=\hbar(\Delta\boldsymbol{k}+\omega_1 \boldsymbol{i})$, and with amplitude $B_{eff}$. The $\mathrm{A_j}$ and $\mathrm{B_j}$ dipolar terms (Eq. \ref{DipolairesA}) are unchanged in the unitary transformation. If they are treated as a time independent perturbation of $\hbar \Delta s_z$+$\hbar \omega_1 s_{x}$+$\hbar \Delta_n I_z$, in a first-order approximation, the $\mathrm{B_j}$ terms have no contribution and thus should be eliminated. If the remaining part of $U_1 \mathcal{H}_{den} U_{1}^{+}$, being time-dependent, is treated as a time-dependent perturbation of $\hbar \Delta s_z+\hbar \omega_1 s_{x}$ +$\hbar \Delta_n I_z$+$\Sigma_{j}f(r_j)A_j$, these other transformed terms have no first-order contribution and should also be eliminated. For instance  a $C_j$ dipolar term contains a $s^{+}I_{jz}$ operator, transformed into $s^{+}I_{jz}$ $e^{i\omega t},$ which induces an electron Zeeman transition, but the energy difference cannot be compensated for by the rapid fluctuations of the dipolar energy, with a far weaker amplitude. In a first-order approximation, one should therefore replace $\mathcal{H}_1$ by the following truncated Hamiltonian $\mathcal{H}_1^{'}$:
\begin{equation}\label{hprime1} 
\mathcal{H}_1^{'}=\hbar \Delta s_{z}+\hbar \omega_1 s_x+\hbar \Delta_n I_z+\mathcal{H}_{den}'
\end{equation}
with
\begin{equation}\label{HdPrime1}
\mathcal{H}_{den}^{'}=\sum_{j}K_js_{z}I_{jz}, \quad K_j=f(r_j)(1-3\cos^2{\xi _j}).
\end{equation}
A second unitary transformation is now made \cite{Redfield1955}, with $U_2=e^{i\Theta( s_y+I_y)}$, where $\Theta$ is the angle between the static field and the effective field $\boldsymbol{B_{eff}}$ acting on spin A:
\begin{equation}\label{DefinitionTHETA}
\cos \Theta =\frac{\Delta }{\Omega _{R}},\ \ \ \sin \Theta =\frac{\omega _1}{\Omega _R} ,
\end{equation}
with $\Omega _{R}=$ $\sqrt{\omega _{1}^{2}+\Delta ^{2}}$ (Rabi frequency for spin A). When $\mid \Delta\mid$  $\lesssim$ $\delta$ ($\delta$: half-width of the unsaturated ESR line), then $\Delta <<\omega _1,$ and $\Theta-(\pi /2)\simeq -\Delta /\omega _1$ (at resonance $\Theta =\pi /2$). $U_2$ is time independent, and hence instead of Eq. (\ref{hamiltonienH1}) one gets:\begin{equation}
\mathcal{H}_2=U_2\mathcal{H}_1' U_{2}^{+}.
\end{equation}
The transform of $\hbar\Delta s_z+\hbar \omega_1s_x$ is $\hbar \Omega _Rs_z$:
 after the Product transformation $U_P$ = $U_2 U_1,$ spin A is submitted to an effective field along $Oz$ with the amplitude $B_{eff}.$ The transform of the remaining part of $\mathcal{H}_1^{'}$ is time-independent and is treated with first-order time-independent theory. $U_2 \hbar \Delta_n I_zU_2^\dagger$ is a sum of two terms. The first one involves a $\cos\Theta I_z$ operator. The second one, with a $\sin \Theta I_x$ operator, has no first-order contribution. The transform $U_2s_zI_{jz}U_2^\dagger$ (cf. Eq. (\ref{HdPrime1})) is a sum of four terms. Only its $\cos ^{2}\Theta s_zI_{jz}$ term does contribute. Since, starting from $\mathcal{H},$ one makes a truncation after each unitary transformation, the final Hamiltonian is not strictly the transform of $\mathcal{H}$ through the product $U_P$, and will be denoted as $\mathcal{H}_{P}',$ the prime recalling these truncations:
\begin{equation}\label{HprimeP}
\mathcal{H}_P'=\hbar \Omega_Rs_z+\hbar \Delta_n I_z \cos \Theta + \sum_jK_js_zI_{jz}\cos^{2}\Theta .
\end{equation}  
A comment about the interpretation of the transformation induced by $U_1$ or $U_2$ is necessary. When a quantum system has an angular momentum $\boldsymbol{l}$, quite generally the transformation induced by $U_1$=
$e^{i\alpha\boldsymbol{l}}$ may be interpreted (cf. Messiah, p. 526-527 of \cite{Messiah1962}) either as a rotation of the reference frame with an angle $\alpha$ (active interpretation), or as a rotation of the system with the opposite angle (passive interpretation). In order to keep a link with results from classical mechanics, it has been usual to use the active interpretation in the field of MR, since its early days, and e.g. to use the so-called rotating frame  \cite{Rabi1954}. We will follow Messiah and adopt the passive interpretation (rotation of the sample). Both interpretations should be seen as corresponding to mental operations (a 10 GHz rotation  of the sample is inconceivable, and the situation would be even worse if trying to rotate the reference frame and any observer tied to it).  And moreover, when  one deals with a spin 1/2 and not with an orbital angular momentum, both interpretations should be taken with a grain of salt, because a spin 1/2 is a spinor, and in a $2 \pi$ rotation a given ket is changed into its opposite (cf. p. 32 of \cite{Slichter1990}).
\section{Transient behaviour of the central spin\label{TransientBehavior}}
\subsection{The reduced density operator\label{SubsectionReducedDensityOp}}
At the chosen time scale, system $\Sigma$ (central A electron spin , $\Sigma_e$, state space  $\mathcal{E}_{e}$, and its $N$ neighbouring B nuclear spins, $\mathrm{\Sigma_n}$, $\mathcal{E}_n$) is effectively coupled only to the static field $B_0\boldsymbol{k}$ and to the oscillating magnetic field $2B_{1}\cos \omega t\boldsymbol{i}.$ The transient behaviour of e.g. the mean value of any component of $\boldsymbol{s}\,$(spin A) is completely defined by the reduced density operator $\rho_{\mathrm{A}}(t)=\mathrm{Tr}_{\mathcal{E}_n}{\rho(t)}$, with $\rho (t)$ defined \cite{Dirac1958} according to:
\begin{equation}\label{RhoIndiceSDet}
\rho(t)=\sum_m p_m \mid m(t)><m(t)\mid,
\end{equation}
with
\[ \mid m(t)>=U(t,0)\mid m>, \]
where $\mid m>$ is one of the $2^{N+1}$ states of an orthonormal basis of $\mathcal{E=E}_e\otimes\mathcal{E}_n,$ $0\leq p_{m}\leq 1$ and $\sum_mp_m=1$, and $U(t,0)$ is the time-evolution operator driving the evolution of $\Sigma$ from its initial mixed or pure state described by:
\begin{equation}
\rho (0)=\sum_mp_{m}\mid m><m\mid .  \label{RhoSde0}
\end{equation}
In the unitary transformation through $U_1$, followed by a truncation of the transformed Hamiltonian, $\mathcal{H}$ is transformed into $\mathcal{H}_1^{'}$, which is time-independent, and $\mid m(t)>$ transformed into $\mid m_1(t)>$, approximated as $\mid m_1'(t)>=$ $e^{-i\mathcal{H}_\mathit{1}'\mathit{t}/ \hbar}$ $\mid m>$. The second unitary transformation, through $U_2$, followed by a truncation of $\mathcal{H}_2$, finally leads to $\mathcal{H}_P'$ (Eq. (\ref{HprimeP})), and $\mid m_1'(t)>$ is transformed into:
\begin{equation}
\mid m_P'(t)>=e^{-i\mathcal{H}_{P}'t/\hbar }\mid m_{P}'(0)>,
\end{equation}
with
\begin{equation}
\mid m_P'(0)>=U_2\mid m>.
\end{equation}
$\rho (t)$ is consequently transformed into $\rho _P(t)=P\rho(t)P^\dagger $, which will be approximated as
\begin{equation}\label{RhoPrimeP(t)}
\rho_P'(t)=e^{-i\mathcal{H}_\mathit{P}'t/\hbar}\rho_P'(0)e^{+i\mathcal{H}_\mathit{P}'t/\hbar} ,
\end{equation}
with
\begin{equation} \label{RhoPrimeP(0)}
\rho _P'(0)=U_2\rho(0)U_2^\dagger .
\end{equation}
In order to presently focus on the transient nutations, instead of using $\mathrm{Tr}\{\rho (t)s_{x}\}$ $=$ $\mathrm{Tr}\{\rho _{P}(t)s_{xP}\}$ (laboratory - tied observer's viewpoint), one should consider $\mathrm{Tr}\{\rho _{P}(t)s_{x}\}$ (sample - tied observer's viewpoint), approximated as $\mathrm{Tr}\{\rho _P'(t)s_{x}\}.$ Since $s_x$ is a component of spin A, it is relevant to introduce the following reduced density operator acting in $\mathcal{E}_e$:
\begin{equation}
\rho _{Pe}^{\prime }(t)=\mathrm{Tr}_{\mathcal{E}_n}\{\rho _{P}^{\prime }(t)\}, \label{OpTraceReduite}
\end{equation}%
 which allows us to write the mean value of $s_x$ as $<s_x>$ = $\mathrm{Tr}_{\mathcal{E}e}\{\rho _{P_e}'(t)s_x\}$, where $\rho _{P_{e}}'(t)$ (cf. Eq. (\ref{RhoPrimeP(t)})) is a function of $\mathcal{H}_P'$. Eq. (\ref{HprimeP}) is now written as
\begin{equation}
\mathcal{H}_{P}'=h_{Ze,eff}+\mathcal{H}_{Pen}'+\mathcal{H}_{\Delta n}, \label{Hprime1PSum4}
\end{equation}%
with $h_{Ze,eff}=\hbar\Omega _Rs_z$, $\mathcal{H}_{Pen}'$ = $\sum_jK_js_zI_{jz}\cos^{2}\Theta$ (which cancels at resonance) and $\mathcal{H}_{\Delta n}$ = $\hbar \Delta_n I_z \cos \Theta$. In Eq. (\ref{Hprime1PSum4}), each term in the sum commutes with the two other terms. This property and the invariance of the trace under a cyclic permutation lead to:
\begin{equation} \label{ContientNewTrace}
\rho_{Pe}'(t)=e^{-ih_{Ze,eff}t/\hbar}\mathrm{X} e^{+ih_{Ze,eff}t/\hbar},
\end{equation}
with
\begin{equation}\label{NewTrace}
\mathrm{X}=\mathrm{Tr}_{\mathcal{E}_n}\{e^{-i\mathcal{H}_{Pen}'t/\hbar}\rho_P'(0) e^{+i\mathcal{H}_{Pen}'t/\hbar}\} .
\end{equation}
The following reduced length $\Re _j,$ reduced dipolar energy $k_j,$ reduced time $\tau $ and effective reduced time $\tau _{eff}$ are now introduced:
\begin{equation} \label{rhojEtkj} 
\Re _j =\frac{r_j}{a}\mbox{, }k_j=\frac{K_{j}}{K_{0en}},\mbox{ \ } \tau _{eff} =\tau\cos^2 \Theta ,
\end{equation}
with\begin{equation}\label{TauEtTaueff} 
\tau =\frac{t}{t_{ref}},\mbox{ }t_{ref}=\frac{\hbar}{K_{0en}}.
\end{equation}
$a$ is a characteristic length to be explicited in each specific case, and the dipolar parameters $K_{0en}$ and $k_j$ are:
\begin{equation}
K_{0en}=\frac{\mu_0}{4\pi }\frac{g_{e}g_{n}\mu _{B}\mu _{N}}{a^{3}},\mbox{  }k_j=\frac{1}{\Re _j^3}(1-3\cos^2\xi _j) . \label{K0eS}
\end{equation}%
In Eq. (\ref{ContientNewTrace}), the partial trace may then be written as:
\begin{equation}
\mathrm{X}=\mathrm{Tr}_{\mathcal{E}_{n}}\{e^{-i\tau _{eff}\sum_{j}k_{j}I_{jz}s_{z}}\rho
_{P}^{\prime }(0)e^{i\tau _{eff}\sum_{j}k_{j}I_{jz}s_{z}}\} .
\label{PartialTrace}
\end{equation}
$\rho (0)$ will be supposed to be a product:%
\begin{equation}
\rho (0)=\rho _{e}(0)\otimes \rho _{n}(0) ,
\end{equation}
e.g. found when initially $\Sigma _{e}$ and $\Sigma _{\mathrm{n}}$ are both at thermal equilibrium, or separately prepared in a pure state.
\subsubsection{$\mathrm{\Sigma_e}$ and $\;\mathrm{\Sigma_n}$ initially at thermal equilibrium} \label{AetLesBjATemperatureT}
One presently starts from
\begin{equation}\label{rhoZerolesdeuxàT} 
\rho(0)=\frac{\exp{[-(\mathcal{H}_Z+\mathcal{H}_{den}')/kT ]}}{\mathrm{Tr}\{\exp{[-(\mathcal{H}_Z+\mathcal{H}_{den}'})/kT]\}},
\end{equation}
with $\mathcal{H}_Z=\mathcal{H}_{Ze}+\mathcal{H}_{Zn}$, where $\mathcal{H}_{Ze}$ and $\mathcal{H}_{Zn}$ were defined at the beginning of Section \ref{SpinsAEtB-Modele}, and $\mathcal{H}_{den}'$ is the secular part of $\mathcal{H}_{den}$.
In conventional CW ESR, the High Temperature Approximation (HTA) $\hbar\omega _0/kT<<1$ is valid even at $T=4$ $K$ (if $\omega_0/2\pi= $ $10$ $GHz$ and $T=4.2$ $K$, then $\hbar \omega _0/kT=0.11$), and it is even more valid for the B nuclear spin neighbours ($\mu _N<<$ $\mu _B$). Both the HTA and the weak dipolar coupling condition are presently assumed. In the first-order expression for $\rho(0)$, one may neglect the correction from $\mathcal{H}_{den}'$ compared with that from $\mathcal{H}_Z$ (cf. Appendix: A), and \cite{Goldman1970}). Then
\begin{equation}\label{RhoInitialGlobal}
\rho(0)\simeq \frac{1}{2^{N+1}}(1-\frac{\mathcal{H}_Z}{kT}),
\end{equation}
In Eq. (\ref{RhoInitialGlobal}), the first term corresponds to infinite temperature, the second one is the Zeeman finite temperature correction. Eq. (\ref{NewTrace}) becomes:
\begin{equation} \label{PartialTraceHauteTemp}
\mathrm{X}=\frac{1}{2^{N+1}}\mathrm{Tr}_{\mathcal{E}_n}\{e^{-i\tau_{eff}\sum_jk_jI_{jz}s_z}U_2
(1-\frac{\mathcal{H}_{Ze}}{kT}-\frac{\mathcal{H}_{Zn}}{kT})U_2^\dagger e^{i\tau_{eff}\sum_jk_jI{jz}s_z}\}.
\end{equation}
The first term gives the contribution $1/2$. The contribution from the third term cancels. One is then left with the Zeeman contribution from the A spin:
\begin{equation}\label{ZeemanSpinAExpress1}
\frac{-\hbar\omega _0}{2^{N+1}kT}\mathrm{Tr}_{\mathcal{E}_n}\{e^{-i\tau _{eff}\sum_jk_jI_{jz}s_{z}}e^{+i\Theta s_y}s_ze^{-i\Theta s_y}e^{i\tau _{eff}\sum_jk_jI_{jz}s_{z}}\}.
\end{equation}
In that trace within $\mathcal{E}_n$, $e^{+i\Theta s_y}s_ze^{-i\Theta s_y}$ is equal to $\cos\Theta s_z-\sin\Theta s_x.$ The contribution from the $\cos\Theta s_z$ term to that partial trace is $2^N s_z\cos \Theta$. The second term, $- \sin \Theta s_x,$ gives rise to the following contribution (cf. Appendix: B)):
\begin{equation}\label{ExpressionDePi}
-2^N \mathbf{\Pi} s_x \sin \Theta,  \qquad \mathbf{\Pi} =\prod_{j=1}^N \cos (\tau _{eff}\frac{k_j}{2}),
\end{equation}
and finally, in $\rho _{Pe}'(t),$ to the appearance of the Rabi frequency operator
\begin{equation}
\widetilde{s}_{x}=e^{-i\Omega _Rts_z}s_xe^{+i\Omega _Rts_z}=s_x\cos\Omega _Rt+s_y\sin \Omega _Rt
\end{equation}
The first-order approximation for the reduced density operator is
\begin{equation}\label{RhoPrimeP(t)HauteTemp}
\rho _{Pe}'(t)\simeq \frac{1}{2}-\frac{\hbar \omega _0}{2kT} \mbox{ }(s_z \cos \Theta -\widetilde{s}_x \mathbf{\Pi} \sin\Theta) .
\end{equation}%
Therefore, $\rho _{Pe}'(t)$ contains a time-independent diagonal part, and a time-modulated off-diagonal part $\widetilde{s}_x  \mathbf{\Pi} \sin\Theta$ which starts with the value $s_x \sin\Theta,$ whatever the strength of the dipolar coupling. The density operator $\rho_{Pe}'(t)~$ drives $<s_i(t)>=\mathrm{Tr}_{\mathcal{E}_e}\{\rho _{Pe}^{\prime }(t)s_{i}\}$, the sample-tied observer mean value of any component of $\boldsymbol{s}$, leading to:
\begin{eqnarray} \label{<s_x>}
<s_x(t)>&=&A\mathbf{\Pi} \sin\Theta\cos\Omega _Rt ,  \\ \label{<s_y>}
<s_y(t)>&=&A \mathbf{\Pi} \sin \Theta \sin \Omega _Rt ,  \\\label{<s_z>}
<s_z(t)>&=&-A\cos\Theta, \mbox{ with }A=\frac{\hbar\omega _0}{4kT}  . 
\end{eqnarray}
Eq. (\ref{<s_x>}) and (\ref{<s_y>}) show that $<s_x(t)>$ and $<s_y(t)>$ oscillate at the Rabi frequency $\Omega _R$ around the effective field, and that the decay of these oscillations  is presently described by the factor $\Pi$. In the context of ESR, the difficulty of a direct observation of these oscillations was stressed in Section \ref{Introduction} and, faced with this difficulty, trying moreover to prepare the resonant spins in some particular state just before the beginning of the microwave pulse should be rather difficult. However, a more general context than ESR does exist, and finding the expression of the reduced density matrix after a preparation of $\mathrm{\Sigma_{e}}$ in a pure state is of interest.
\subsubsection{Initially, $\mathrm{\Sigma_e \;} $ in state $ \mid->$ and $\mathrm{\Sigma_n\;}$ at T}\label{Adansetat-etLes BaT}
An initial state with the central electron spin in its $\mid ->$ state and the B nuclear spins at thermal equilibrium is therefore now assumed. In the HTA for the B spins, the initial density operator is then
\begin{equation}
\rho (0)\ =\frac{1}{2^{N+1}}(1-2s_z)\otimes(1-\frac{\mathcal{H}_{Zn}}{kT}) .
\end{equation}%
The $\mathcal{H}_{Zn}$ contribution to $\rho _{Pe}'(t)$ again cancels out. The contribution of the $s_z\sum_{j}I_{jz}$ term can be neglected compared with that of the central spin Zeeman $s_z$ term ($\hbar\omega_{0n}<<kT$ even at $1 K$, and this term moreover implies a factor which, when $\mathbf{\Pi} $ has a value near $1,$ is itself lower than $1$). This leads to:
\begin{equation}\label{Rho'P(t)AEtat-BaT}
\rho _{Pe}'(t)\simeq \frac{1}{2}-s_z \cos \Theta +\widetilde{s}_x \mathbf{\Pi} \sin\Theta.  
\end{equation}%
The $\hbar\omega _{0}/2kT$ factor which affected both $s_z$ and $\widetilde{s}_x$ in (\ref{RhoPrimeP(t)HauteTemp}) is absent in (\ref 
{Rho'P(t)AEtat-BaT}), but the off-diagonal elements are again dominated by the factor $ \mathbf{\Pi} ,$ studied in detail in Subsection \ref{SubsectionDecoherence}.\\
The situation with, initially, A in state $\mid ->$ and the nuclear spins in state $\mid +>$ would necessitate either a temperature far lower than $ 1 K$ or some specific manipulation of the nuclear spins, and thus will not be examined.
\subsection{Decay of the TN through dipolar coupling\label{SubsectionDecoherence}}
At $t=0, \mathbf{\Pi}=1.$ In this subsection and in the next one, in order to appreciate the time dependence of $ \mathbf{\Pi}$,  the B spins occupy the nodes of a simple cubic lattice, with a unit cell of length $a$ and a four-fold symmetry axis parallel to the static field $B_{0}\boldsymbol{k}$. Spin A, substituted to a B spin, occupies a node taken as the coordinate origin. In Subection \ref{SubSectionDiluteSamples}, 
the B sites are occupied randomly by a B spin.

In $ \mathbf{\Pi}$ defined in Eq. (\ref{ExpressionDePi}), the effective time is $\tau _{eff}\ =\cos^{2}\Theta .K_{0en}t/\hbar$ (cf. Eq. (\ref{TauEtTaueff})). When dipolar coupling is absent, $K_{0en}=0$ and as time increases $ \mathbf{\Pi}$ keeps equal to 1 (cf. Eq. (\ref {RhoPrimeP(t)HauteTemp})). The presence of the dipolar coupling is described in Eq. (\ref{RhoPrimeP(t)HauteTemp}) through the product $ \mathbf{\Pi}$, decreasing from $1$ towards $0$ as time goes on, as shown below. 

As time increases from $t=0,$ the\ first $ \mathbf{\Pi}$ cancellation is produced by\ the maximum possible $\mid k_{j}\mid $ value, and occurs at the smallest instant for which
\begin{equation}
\frac{\tau _{eff}\ \mid k_{j}\mid _{\max }}{2}=\frac{\pi }{2}.
\end{equation}%
Since $k_{j}=(1-3\cos ^2\xi _j)/\Re _j^3,$ this smallest $\tau_{eff} $ value is produced by the B neighbours for which both $\cos ^{2}\xi
_j=1$ and $\Re _j=1,$ i.e. from the two neighbours with respective coordinates $(0,0,$ $1)$ and $(0,0,$ $-1)$,~and therefore for which $%
k_{j}=-2 $. This first $ \mathbf{\Pi}=0$ value thus occurs when $\tau _{eff}=\pi /2.$ More generally, these two nearest neighbours (n.n.) produce a zero value for $ \mathbf{\Pi}$ when:
\begin{equation}
\tau _{eff}=(2m+1)\frac{\pi }{2} \mbox{ for } m=0,1,2....
\end{equation}
Since the four other n.n., with coordinates $(\pm 1,0,0)$ and $(0,\pm 1,0)$, have $k_j=1,$ the effective time values corresponding to a cancellation of $ \mathbf{\Pi}$ are $\tau _{eff}=(2m+1)\pi .$ 
Among the twelve next nearest neighbours, situated at $\sqrt{2}$ from spin A, four of them, $((\pm 1,\pm 1,0))$, cancel $ \mathbf{\Pi}$ for $\tau _{eff}=2\sqrt{2}(2m+1)\pi ,$  while the eight others cancel $ \mathbf{\Pi}$ for $\tau _{eff}=4\sqrt{2}(2m+1)\pi .$    Therefore, the smallest $\tau _{eff}$ value for which $ \mathbf{\Pi}=0$ is $\tau _{eff}=\pi /2,$ and when $\tau _{eff}$ increases: 1) a given choice of neighbours is responsible for a periodic appearance of $ \mathbf{\Pi}$ cancellation, 2) the different periods are not all commensurate. 
This behaviour establishes that $ \mathbf{\Pi}$  is not periodic, and that on the contrary it decreases from its initial value $\mathbf{\Pi}=1$ towards $\mathbf{\Pi}=0$, but not strictly in a monotonous way. As a result of symmetry, the number of occurrences of each $k_j$ value is even, and therefore necessarily $ \mathbf{\Pi} \geq 0.$
$ \mathbf{\Pi}$ has been calculated for B neighbours with coordinates $(i,j,k)$ chosen to obey $\mid i\mid \leq i_{\max }$, $\mid j\mid \leq j_{\max}$, $\mid k\mid \leq k_{\max}$, and taking a Common Maximum value for $\mid i\mid ,$  $\mid j\mid $, $\mid k\mid ,$ denoted as $CM$. If $CM=1,$ the number of considered neighbours, $N^{\prime}=(2.CM+1)^{3}-1,$ is $26$ (the $6$ n.n, $12$ n.n.n, and $8$ third nearest neighbours). If $CM=5$, $N^{\prime }=$ 1330. Calculations indicate that at any time $\mid (\mathbf{\Pi}_{CM=1}-\mathbf{\Pi}_{CM=5}))/\mathbf{\Pi}_{CM=1}\mid <0.02$, and therefore, from now on, the value $CM=1$ (26 neighbours) will be used. From Fig. 1, it appears that, for a sample-tied observer, the effective time for the nutations to disappear, which we denote as $\tau _{eff,DC}$ (effective Dipolar Coherence time) is such that:
\begin{equation} \label{TaueffDec}
\tau _{eff,DC}=\frac{\pi }{2}.
\end{equation}
The $<s_x(t)>/<s_x(0)>$ ratio may be written (cf. Eq. (\ref{<s_x>})) as:
\begin{equation} \label{sx(t)/sx(0)} 
\frac{<s_x(t)>}{<s_x(0)>}=\mathbf{\Pi} \cos \Omega _{R,eff} \tau _{eff}, 
\end{equation}
with
\begin{equation}
\Omega _{R,eff}=\frac{\hbar\Omega _R}{\cos^{2}\Theta K_{0en}}.
\end{equation}
The time behaviour of the ratio defined in Eq. (\ref{sx(t)/sx(0)}) is given in Fig. 2A, 2B and 2C, for the respective values $\Omega _{R,eff}=100,\:10$ and $1$, as a function of $\tau_{eff}/\pi.$ The true coherence time corresponding to $\tau _{eff,DR}$ is (cf. Eq. (\ref{TauEtTaueff})):
\begin{eqnarray}
t_{DC}&=&\frac{\pi\hbar}{2K_{0en}\cos^{2} \Theta} \label{TempsvraiDec1}\\
&\simeq& \frac{\pi\hbar}{2K_{0en}}\frac{\omega_1^2}{\Delta^2}, \qquad\Delta<<\omega_1.\label{TempsvraiDec2}
\end{eqnarray}\\
Equation (\ref{TempsvraiDec1}) indicates that at exact resonance the coherence time is infinite, i.e. at resonance the dipolar coupling does not produce decoherence, which then is produced by the SL coupling, neglected in our approach. A link will now be established between $t_{DC}$ and the shift from the exact resonance of spin A in a CW ESR experiment in linear regime.
\vspace{1 mm}\\
\hspace{-14 mm}\scalebox{0.25}{\includegraphics{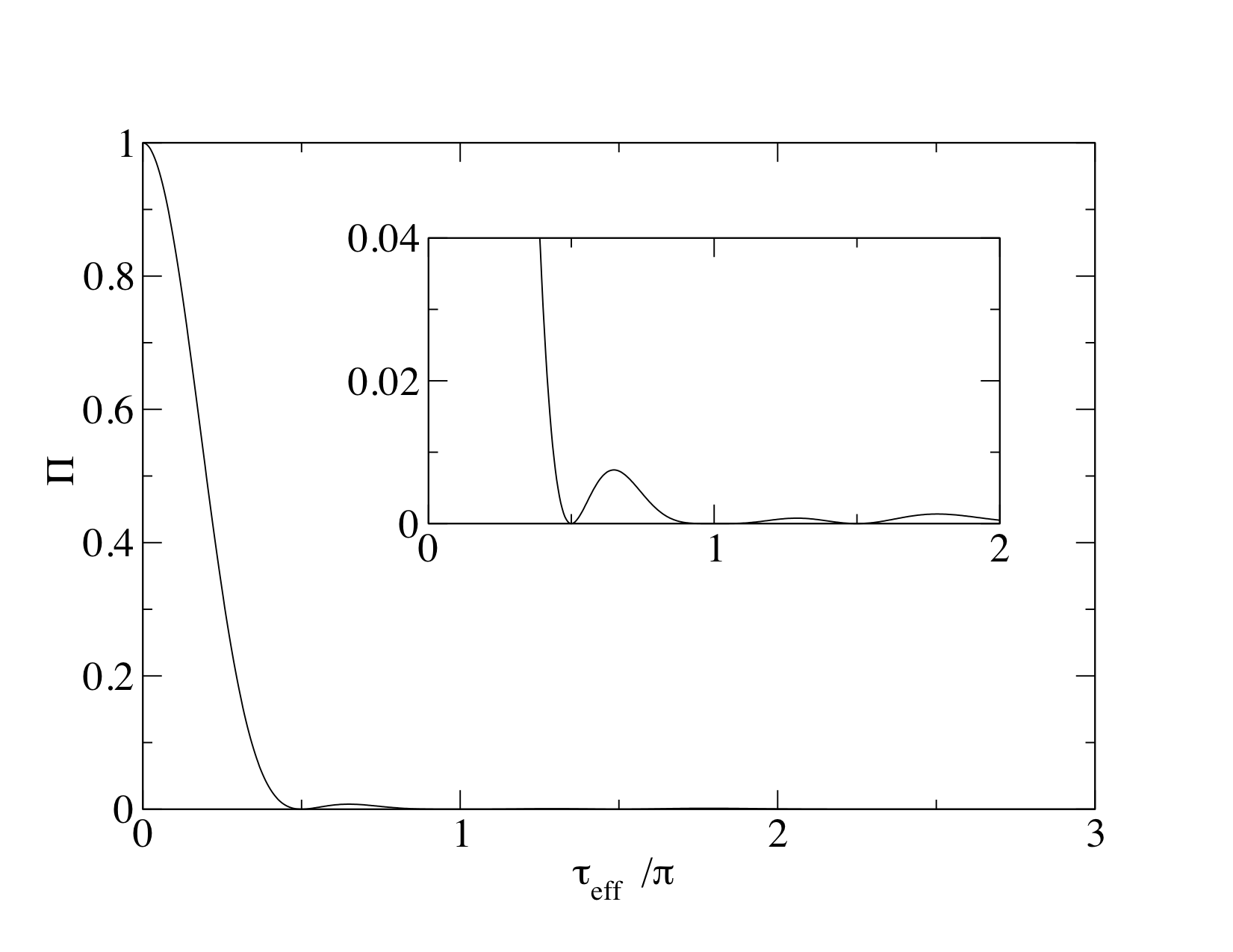}}\\

\vspace{1.5mm} \hspace{-5 mm} Fig. 1 Time variation of $ \mathbf{\Pi}$. Inset: close-up view. 

\vspace{-2mm}

\vspace{-61 mm}
\hspace{91 mm}\scalebox{0.25}{\includegraphics{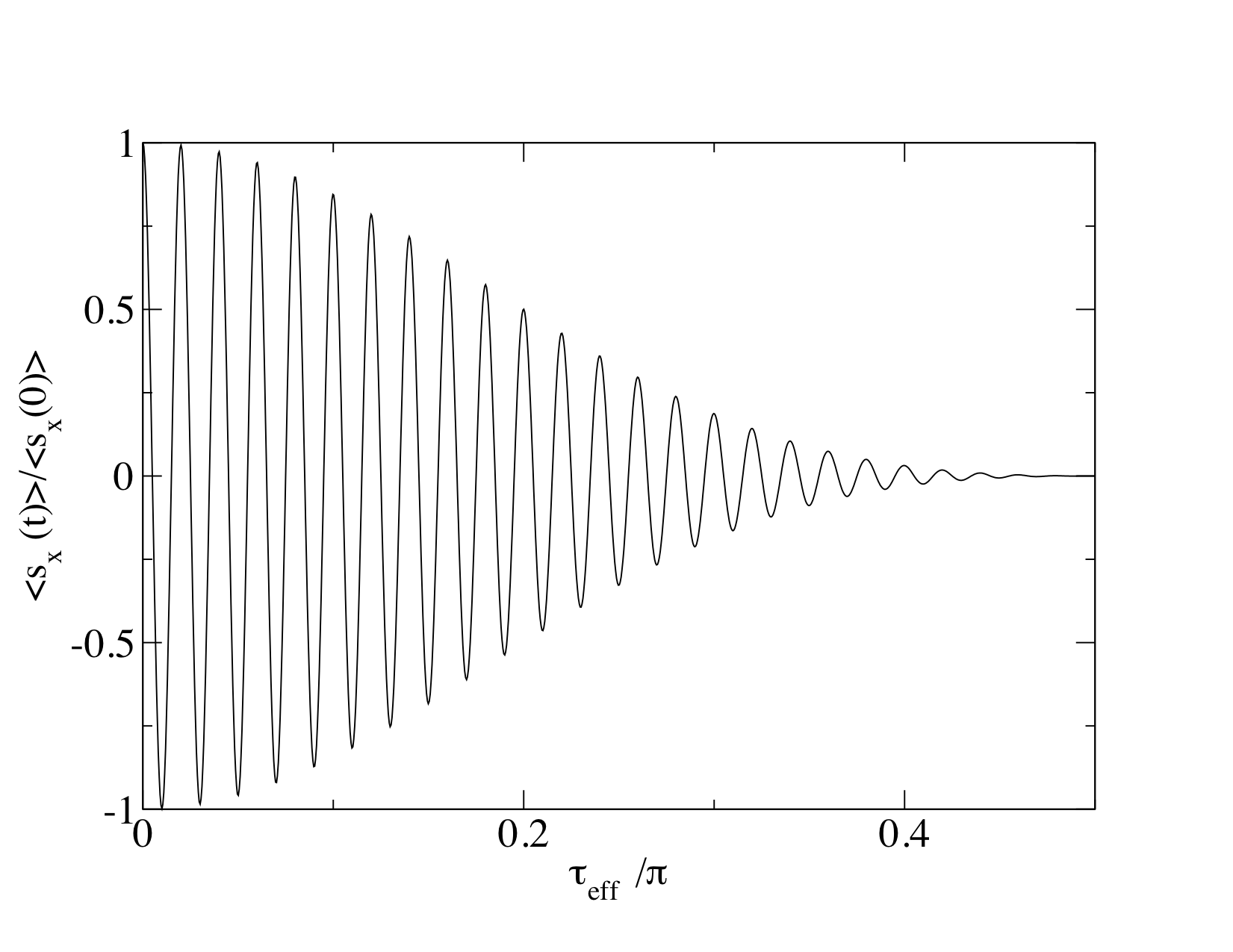}}\\

\hspace{+89 mm}
\vspace{-1 mm}
Fig. 2A Time variation of $\frac{<s_x(t)>}{<s_x(0)>}\mbox{ if } \Omega _{R,eff}=100 $.\\

\hspace{-3mm}\scalebox{0.25}{\includegraphics{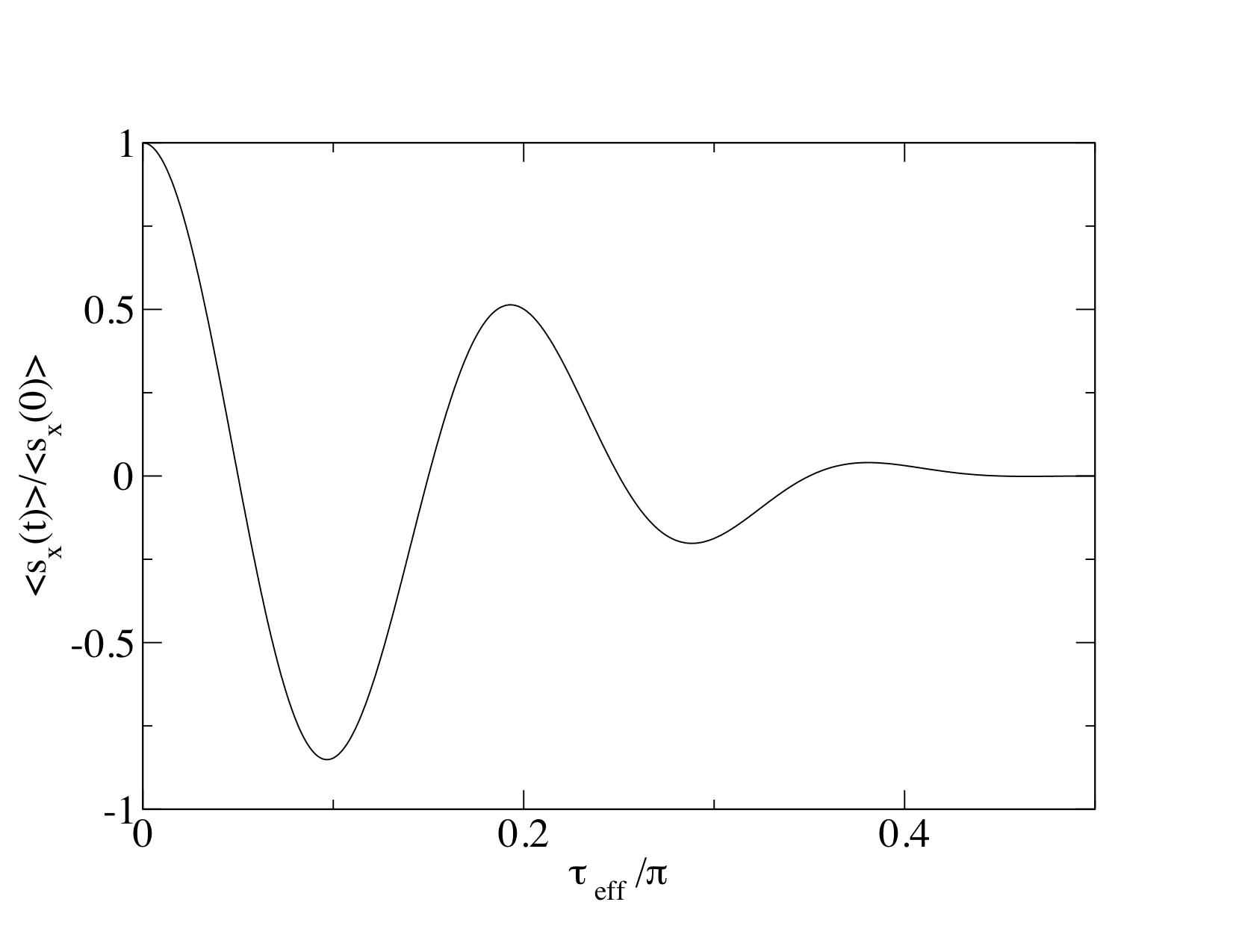}}

\hspace{-6mm} Fig. 2B Time variation of $\frac{<s_x(t)>}{<s_x(0)>}$ \mbox{ if } $\Omega _{R,eff}=10. $ \\
\vspace{11 mm}
  
\vspace{-75 mm}
\hspace{91 mm}\scalebox{0.25}{\includegraphics{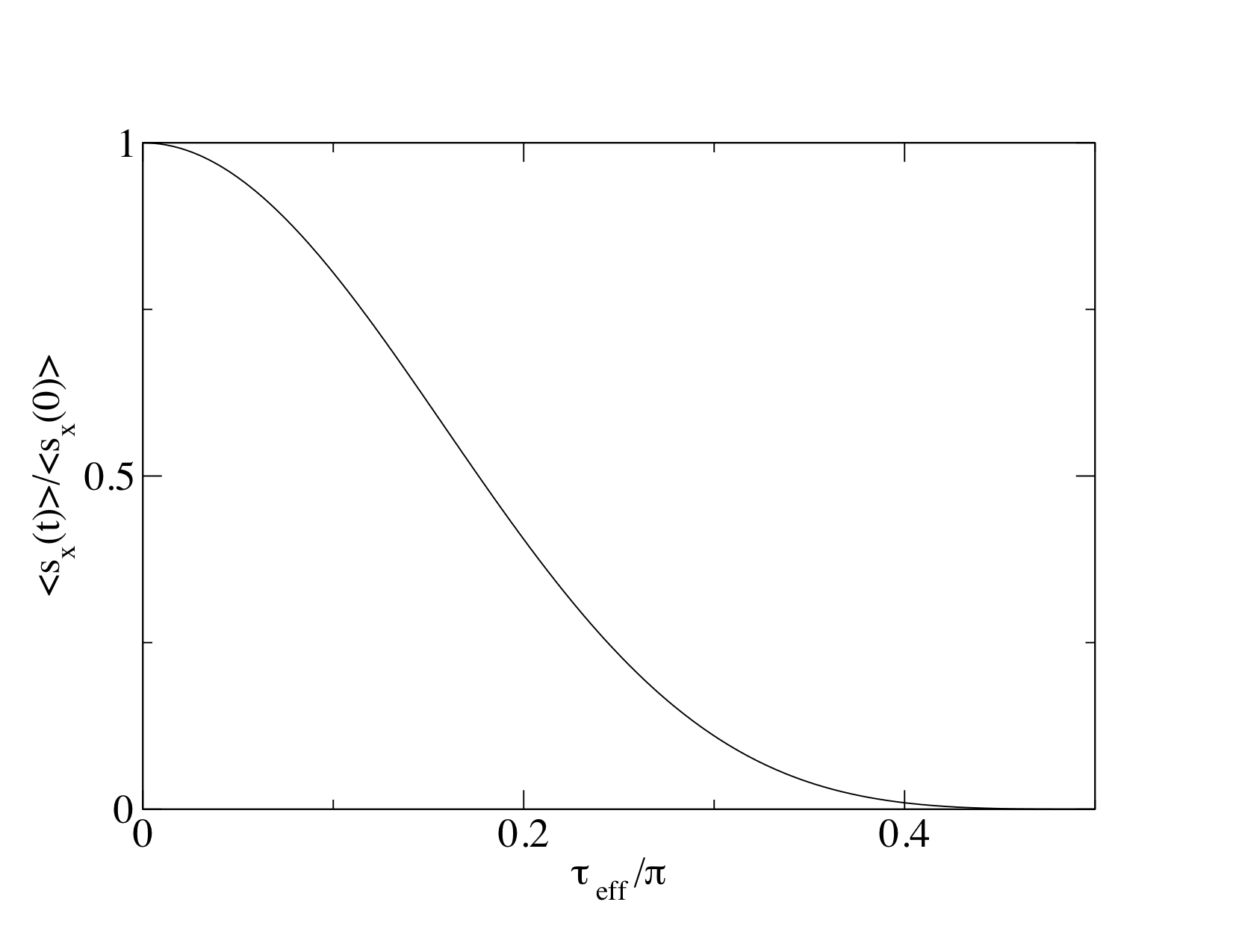}}\\
\vspace{-1 mm}
\hspace{+93 mm} 
Fig. 2C Time variation of $\frac{<s_x(t)>}{<s_x(0)>}\mbox{ if }$ $\Omega _{R,eff}=1$. \\

\vspace{+2 mm}
\subsection{Coherence time and frequency shift}\label{SubSectionFrequencyShift}
In a transient ESR experiment aimed at measuring $t_{DC}$, one should avoid being at exact resonance, since the dipolar coupling then does not produce any decoherence. One should obviously avoid being completely outside resonance, and choosing $\Delta \simeq\delta ,$ the half-width of the absorption unsaturated steady-state ESR line, seems a good balance. One has then to find the theoretical expression for $t_{DC}$ when $\Delta =\delta ,~$\ which imposes a theoretical determination of $\delta .$\ As mentioned in Section \ref{SpinsAEtB-Modele}, the Van Vleck condition is then fulfilled, whereas the weak dipolar coupling condition was supposed to be fulfilled in the transient experiments for which $t_{DC}$ is meaningful. In the following $\delta $ determination, one has therefore to adapt well-established general results for the usual steady-state unsaturated ESR spectrum to the present system, with an electron A spin and its non-resonant neighbouring B nuclear spins.\\
Substantial efforts have been achieved since more than twenty years in order to detect a single spin, particularly with optically detected MR (ODMR), with MR force microscopy (MRFM) (c.f. e.g. \cite{Kohler1993, Gruber1997, Rugar2004}) and, more recently, at low temperature (0.5 K) with Spin Excitation Spectroscopy (cf. \cite{Otte2008} and references therein). These approaches are still under development. The sensitivity of conventional ESR spectrometers does not allow them to detect a microscopic number of spins. Therefore, instead of a single central A spin, one should presently think of a collection of electron A spins, distant enough from each other to have a negligible mutual dipolar coupling, and each surrounded by B nuclear spins. The size of a {Central A - neighbouring B spins\} cluster is high enough to present negligible surface effects and negligible coupling with adjacent clusters. In the discussion about the line width, a single cluster may be considered.\\
The unsaturated dipolar line shape and line width in solids when all the lattice sites are occupied by a spin were explained in \cite{VanVleck1948} with the method of moments, either with a single spin species or, as presently, with two ingredients A and B, A being resonating and B out of resonance. In the discussion to follow, we will refer to \cite{Abragam1961} (Chap. IV), adapting its notations when necessary. In discussions about line widths, a Gaussian or Lorentzian profile is often used. When the line is Gaussian, once its second moment $M_2$ is known, the theoretical half-width is also known: $\delta =$ $1.18\sqrt{M_{2}},$ and the ratio $M_4/M_2^2$ ($M_4:$ fourth moment) is equal to $3$. In order to appreciate the importance of the shift from a Gaussian shape, it is common practice to calculate the ratio $M_4/M_2^2,$ far greater than $3$ for a cut-off Lorentzian. The second moment, defined as in \cite{Abragam1961}, but with the present notations, and with $\gamma _e\hbar=-g_e\mu _B$ and $\gamma _n\hbar=g_n\mu _N$, is denoted $M_2'$ to stress the fact that two ingredients are present (cf.\cite{Abragam1961}, Chap. IV, Section III). $M_2'$ presently satisfies:
\begin{equation}\label{SecondMoment}
\hbar^2M_2'=
-\frac{\mathrm{Tr}\{[\mathcal{H}_{den}',s_x]^2\}}{\mathrm{Tr}\{s_x^2\}},
\end{equation}
since $\mathcal{H}_{den}'$ is presently the secular part of $\mathcal{H}_{den}$. The traces should be calculated within 
$\mathcal{E=E}_e\otimes\mathcal{E}_n.$ One gets:

\begin{equation} \label{SecondMomentf=1}
\hbar^2M_2'=\frac{1}{3}K_{0en}^2I(I+1)\sum_jk_j^{2} \simeq 3.34K_{0en}^2,
\end{equation}
since $=1/2$ and since, for the chosen case (simple cubic lattice, static field along a four-fold symmetry axis), $\sum_{j}k_{j}^{2}$ $\simeq $ $13.35.$
$M_4'$ satisfies 
\begin{equation}\label{QuatriemeMomentf=1}
\hbar^4 M_4'=
\frac{\mathrm{Tr}\{[\mathcal{H}_{den}',[\mathcal{H}_{den}',s_x]]^2\}}{\mathrm{Tr}\{s_x^2\}}. 
\end{equation}
The calculation of $M_4'$, generally quite cumbersome, is presently lighter, because each cluster is assumed uncoupled to the other clusters, and moreover a given cluster contains a single resonant spin. One gets (cf. Appendix: C)):
\begin{equation} \label{4emeMomentAvecSommesf=1}
\hbar^{4}M_4'=\frac{K_{0en}^4}{16}[\sum_j k_j^4+3\sum_{j,l\neq{j}}k_j^2 k_l^2].
\end{equation}
The summations involve the B sites around A. Presently, $\sum_{j}k_{j}^{4}$ $\simeq≃36.1$, and $\sum_{j,l\neq j}k_{j}^{2}k_{l}^{2}$ $\simeq142.1$, and therefore:
\begin{equation}
\hbar^4M_4'=28.9K_{0en}^4.
\end{equation}
The theoretical ratio $M_{4}^{\prime }/M_{2}^{^{\prime }2}$ is therefore equal to $2.6$, close to its value for a Gaussian profile with half-width $\delta =$ $1.18\sqrt{M_{2}^{\prime }}=$ $2.16K_{0en}/\hbar,$ and the coherence time for $\Delta =\delta $ is therefore (cf. Eq. (\ref{TempsvraiDec2})) approximately:
\begin{equation}\label{tdrDelta=delta}
(t_{DC})_{\Delta =\delta}=\frac{3.39 B_{1}^2}{\mid \gamma _{e}\mid \delta _{B}^{3}},
\end{equation}
where $\delta _B=\delta /\mid \gamma _e\mid $ is the half-width expressed in magnetic field units. \\
If\ $a=3$ \AA , $g_e=2,$ $g_{n}=5.586$ (proton), then $K_{0en}=1.94\times 10^{-27}$ $J$ and $\delta _B$ $=$ $ 0.23$ $mT.$ If moreover $B_{1}=1 $ $mT$, then $(t_{DC})_{\Delta =\delta }\simeq 1.7$ $\mu s$. If $g_{n}=2.261$ (as for the $^{31}P$ stable isotope, with nuclear spin $I=1/2$ ),  and again $g_{e}=2$\ and $B_{1}=1$ $mT$, then $(t_{DC})_{\Delta=\delta }\simeq 25,6$ $\mu s$.\\
These two numerical examples suggest that, for a simple cubic lattice with $a=3$ \AA , with a central electron spin and nuclear neighbours and for $B_1 =10^{-3} T,$ $(t_{DC})_{\Delta =\delta }$ should be in the $1-50 $ $\mu s$ range. There remains to examine how these results should be modified in dilute samples.
\subsection{Coherence time $t_{DC}$ in dilute samples} \label{SubSectionDiluteSamples}
The B sites are now assumed to be ocupied at random, each one with the same probability $f,$ by a B nuclear spin. The other previous assumptions, including an electron A spin, are kept. One must first establish how the calculation which led to Eq. (\ref{TaueffDec}) should be modified when $f<1.$ In Eq. (\ref{ZeemanSpinAExpress1}), the arguments in the exponential operators should now be written $\mp i\tau _{eff}$ $\times$ $\sum_j^{\prime}k_{j}I_{jz}s_{z},$ where the prime stresses that the sum is made over the B spins, to distinguish it from a summation over the B sites. It is simpler to replace this sum by\ the sum $\mp i\tau_{eff}f\sum_{j}k_{j}I_{jz}s_{z}$ made over the sites. In dilute samples, Eqs. (\ref{TaueffDec}) and (\ref{TempsvraiDec1}) are therefore replaced by:

\begin{equation}
f\tau _{eff,DC}=\frac{\pi }{2},\ \ \ t_{DC}=\frac{1}{f}\frac{\pi \hbar}{2K_{0en}\cos^{2}\Theta} .
\end{equation}
One has now to relate the frequency shift $\Delta $ to the half-width 
 $\delta $ of the unsaturated ESR line when the broadening comes from 
the A spin - non-resonant B spins dipolar coupling, using again the method of moments, adapted to the dilute sample case \cite{Kittel1953} . 
Then, in the calculation of the fourth moment, the sums $\sum_{i}^{\prime }$ $\ $and $\sum_{i,j>i}^{\prime }$ over the spins are replaced by the quantities $f\sum_{i}$ and $f^{2}\sum_{i,j>i},$ with summations over the sites. Eqs. (\ref{SecondMomentf=1}) and (\ref{4emeMomentAvecSommesf=1}) are replaced by
\begin{eqnarray}
\hbar^{2}M_{2}^{\prime } &=&3.34fK_{0en}^{2}, \label{MPrime2fegal1} \\
\hbar^{4}M_{4}^{\prime }&=&\frac{K_{0en}^{4}}{16}(f\sum_{i}k_{j}^{4}+3f^{2}\sum_{j,l\neq j}k_{j}^{2}k_{l}^{2})==2.26fK_{0en}^{4}(1+11.8f).\label{MPrime4fegal1} 
\end{eqnarray}
The $M_{4}^{\prime }/M_{2}^{\prime 2}$ ratio then satisfies%
\begin{equation}
\frac{M_{4}^{\prime }}{M_{2}^{\prime 2}}=0.20[\frac{1}{f}+11.8].
\end{equation}%

If $f=0.05$ then $M_{4}^{\prime }/M_{2}^{\prime 2}=6.4,$ and this ratio increases if $f$ decreases. Therefore, if \ $f\leq 0.05,$ then 
$M_{4}^{\prime }/M_{2}^{\prime 2}\geq 6.$ If $ f \lesssim 0.05,$ the profile of the (unsaturated ESR) line is\ therefore far from Gaussian, and may be approximated with a cut-off Lorentzian profile, and a half-width $\delta $ satisfying (cf. \cite{Abragam1961}):
\begin{equation} \label{tDRDilue} 
\delta =\frac{\pi }{2\sqrt{3}}\frac{M_{2}^{\prime 3/2}}{\sqrt{M_{4}^{\prime }%
}}=3.68\frac{f}{\sqrt{1+11.8f}}\frac{K_{0en}}{\hbar}.
\end{equation}
Making $f=1$ in Eq. (\ref{tDRDilue}) is forbidden, as Eq. (\ref{tDRDilue} ) was obtained assuming  $ f \lesssim 0.05.$

If $a=3$ \AA , $f=10^{-2},$ $g_e=2,$ and if the B neighbours are protons ($g_n=5.586)$, then $\delta_B=$ $3.6\times 10^{-6}$\ $T.$ And, if $B_1=10^{-4}$ $T,$ one gets $(t_{DC})_{\Delta =\delta }=6.6$ $ms.$
\section{Discussion} \label{SectionDiscussion}
Results of Section \ref{TransientBehavior} upon the coherence time $t_{DC}$ are reminiscent of some well-known results obtained in quite different MR contexts. First, our result that at resonance ($\Delta =0$) the dipolar coupling does not affect the TN (cf. (Eq. (\ref{TempsvraiDec2})) is reminiscent of the well-known inefficiency of the field inhomogeneity at resonance in liquid state NMR (cf. Fig III.3 of \cite{Abragam1961}). And secondly, Eq. (\ref{tdrDelta=delta}) for  $t_{DC}$ shares some formal similarity with the expression of a time parameter $T_B$ introduced by Torrey \cite{Torrey1949, Abragam1961} in non-viscous liquids investigated by NMR: there, the molecular motions nearly average out the dipolar coupling and when s=1/2 the Bloch equations are valid  \cite{Bloch1956, Abragam1961}. The inhomogeneity of the static magnetic field may then contribute to a broadening of the resonance line. And if a Gaussian distribution in $B_0$ is assumed, with full width $\Delta B_0$, in conditions allowing an observation of the TN, with $B_1$ $\gg$ $\Delta B_0$, at resonance the amplitude of the signal obtained after integration over $B_0$  does not strictly decay according to an $\exp (-t/T_2)$ law when $T_1 \gg T_2$. This amplitude presents a correcting factor $(1+t/T_{B})^{-1/4}$, with $T_B=$ $5.6B_1/\gamma(\Delta B_0^{2})$. The times $(t_{DC})_{\Delta =\delta}$ and $T_{B}$ both increase with $B_1$, and decrease with $\delta_B$ and $\Delta B_0$ respectively. This formal similarity in different contexts reflects the common presence of a competition between the oscillating field $B_1,$ responsible for coherence, and the source of decoherence (neighbouring B spins or field inhomogeneity). However, there is no reason why the analytical expressions should be identical in these different situations, and it is therefore not surprising to find that $(t_{DC})_{\Delta =\delta}$  $\propto B_1^2$ while $T_{B}$  $\propto B_1$, and that $(t_{DC})_{\Delta =\delta}$  $\propto \delta_B^{-3}$ while  $T_B$ $\propto \Delta B_0^{-2}$. \\
In \cite{DiVincenzo1995}, the state of the Not gate discussed by DiVincenzo was changed by a pulse of magnetic field with a well-chosen duration. 
In that context, any magnetic dipolar coupling of the electron spin to neighbouring nuclear spins would have undesired effects.  The present results show that if the gate operates near resonance, this perturbation should be negligible in a first approximation.\\
It seems natural to hope that the above-mentioned qualitative similarity be also found with the TN of paramagnetic defects with nuclear neighbours ($I=5/2$) studied in \cite{Boscaino1993}, or without nuclear neighbours in \cite{Agnello1999}. But in \cite{Boscaino1993} and \cite{Agnello1999} Boscaino \textit{et al} on the contrary found that the duration of the nutations decreased with $B_1$.  This shortening is unexpected, since the nutations are created by the pulse of oscillating field. Then, increasing the amplitude of the pulse should rather favour their persistence.  In \cite{Boscaino1993}, the sample was submitted to a 2.6 mT transient oscillating magnetic field created with a traveling wave tube (TWT) amplifier. This shortening was first interpreted \cite{Boscaino1993} as a result of fluctuations of the oscillating field inducing the TN.  This seems plausible, as the TWT is known to be a noisy device, but that interpretation also means that the shortening is an artefact possibly hiding a $t_{DC}$ increase with $B_{1}.$  Another interpretation, where the fluctuations are created by the spins, was proposed later \cite{Agnello1999}. This mechanism cannot be effective in the situation considered in this paper, since it implies presently forbidden flip-flops between the central electron spin and its neighbouring B nuclear spins.\\
The Bloch equations \cite{Bloch1946, Bloch1956}, describing the time behaviour of the magnetization $\boldsymbol{M}$, usually written in the rotating frame, take the form:
\begin{equation}\label{BlochRepereTournant}
\frac{d\boldsymbol{M}}{dt}=\gamma (\boldsymbol{M}\wedge \boldsymbol{B_{eff}})-\frac{M_x \boldsymbol{i}+M_y \boldsymbol{j}}{T_2} -\frac{(M_z-M_0)\boldsymbol{k}}{T_1}.
\end{equation}%
They were used in the analysis of the first forced transient experiments displaying TN, in NMR \cite{Torrey1949} and in ESR \cite{Atkins1974}, in liquids, but cannot explain the present decay of the nutations. First, while they could explain NMR lines in non-visquous liquids, they have been known for a long time to fail in the presence of dipolar ESR lines in insulating solids, e.g. predicting an unsaturated Lorentzian line shape \cite{Bloch1946} while dipolar lines are known with a Gaussian profile \cite{Bleaney1950, Abragam1971}. Secondly, the Bloch description is valid for homogeneous MR lines, whereas the ESR line for the present collection of independent clusters should be inhomogeneous (cf. below). Thirdly, even when the Bloch equations can describe an unsaturated MR line, they fail in strong saturation, for thermodynamical reasons first identified by Redfield \cite{Redfield1955} (cf. Section \ref{SpinsAEtB-Modele}), who then successfully introduced a Spin Temperature (ST) in the rotating frame, an explanation valid in steady state and strong saturation, but not claiming to describe the transient regime presently examined, which begins just when the strong oscillating field starts on. Fourthly, vector form (\ref{BlochRepereTournant}) for the Bloch equations keeps true in the rotated-reoriented frame (or, in the interpretation used in this paper, in the rotated-reorientated sample). The present collection of independent resonant electron A spins creates a magnetization proportional to the mean value $<\boldsymbol{s}>=\mathrm{Tr}_{\mathcal{E}_e}\{\rho_{Pe}'(t)\boldsymbol{s}\}.$ The time derivatives of the corresponding mean values $<s_i>$ (cf. Eqs. (\ref{<s_x>}), (\ref{<s_y>}) and (\ref{<s_z>})) could again be formally written as in Eq. (\ref{BlochRepereTournant}), with an infinite $T_1$, and $1/T_2$ replaced by $-d(\ln \mid \Pi \mid )/dt$, but this would be artificial, since this last quantity is time-dependent. It is simpler not to stick to the Bloch equations, or to tentative modified Bloch equations (cf. e.g. \cite{Shakhmuratov1997} and, with opposite assumptions, \cite{Asadullina2001}), but to use $<s_x(t)>,\ <s_y(t)>$ and $<s_z(t)>$  directly.\\
The difficulty of observing the TN directly was stressed from the beginning of Section \ref{Introduction}. Hereafter, additional complications are presented through the consideration of two specific paramagnetic systems. The first one is concentrated $\mathrm{ZnS:Mn^{2+}}$. If the $\mathrm{Mn^{2+}}$ concentration is $\mathrm{c=0.2}$, the 9 GHz ESR spectrum has a single line, with a $9.5\:\mathrm{\:mT}$ linewidth \cite{Gayda1972}. But at high $\mathrm{Mn^{2+}}$ dilution, the ESR spectrum has six  equidistant hyperfine packets ($I=5/2$) of five fine structure (s=5/2) lines, and the width of the whole spectrum is roughly 50 mT. The simplicity of the easily observable spectrum when $c \gtrsim  10^{-1}$ is misleading: $s=5/2$ and not $1/2$, and the single exchange-narrowed line reflects a competition between hyperfine field and exchange, the dipolar coupling being then negligible. Stable organic free radicals (s=1/2) used in ESR as standards (DPPH, BDPA) are the second system. At 9 GHz undiluted DPPH presents a single narrow Lorentzian ESR absorption line. But when DDPH is highly diluted, a hyperfine structure is found. In the undiluted sample, the hyperfine field is negligible compared with the dipolar field, and the linewidth is exchange-narrowed, as the effect of the dipolar field is reduced by the exchange coupling \cite{Bloembergen1954, Misra1971}.  BDPA \cite{Koelsch1957} is also used for Dynamical Nuclear Polarization (DNP). The ESR spectrum of BDPA diluted in mineral oil presents a hyperfine structure \cite{Dalal1974}. BDPA is available in crystalline form and its ESR spectrum then consists of a single narrow line, as a result of exchange-narrowing \cite{Burgess1962, Dalal1974}. One should then e.g. keep a distance from the results for BDPA in \cite{DeRaedt2012}, as the experimental results were indirect (FID), and the simulations ignored exchange coupling and exchange-narrowing.

This discussion will end with a few more thermodynamic considerations. The present behaviour, discussed in Section \ref{UseandAbuse}, may be expressed in the language of reservoirs. In MR, when two parts of the Hamiltonian commute, each may have its own temperature \cite{VanVleck1957} and is then called a reservoir, a concept introduced by Bloembergen and Wang \cite{Bloembergen1954} (the exchange reservoir, used e.g. later in ESR in the field of spin-glasses \cite{Deville1985}) and generalised by Provotorov (dipolar reservoir, cf. \cite{Goldman1970}). During the nutation step, the A Zeeman and the B dipolar reservoirs do not exchange energy. Equation (\ref{RhoPrimeP(t)HauteTemp}) shows that $\rho_{Pe}'(t)$ possesses non-diagonal elements when the oscillating field starts on and for a duration roughly equal to $(t_{DC})_{\Delta=\delta}$. This confirms that an observer tied to the rotating-reorientated sample should reject any ST assumption during that transient period.\\
At the time scale of the TN, the coupling of an A spin either to the lattice or to the other A spins is negligible, and the $s_zS_{jz}$ dipolar terms shift the resonance of these A spins, creating spin packets \cite{Portis1953}. The resonance line is strictly inhomogeneous. On the contrary, a MR line is said to be purely homogeneous if its width only comes from processes shortening the life of the spin levels involved in the MR \cite{Abragam1971}.
 In a first approximation, spin packets centred on different frequencies are independent. In a higher order approximation, the $s^{\pm} I_{j}^{\mp }$ terms would allow energy exchanges between these spin packets, i.e. a spectral diffusion \cite{Portis1956, Mims1961} through dipolar coupling (not to be confused with the spatial diffusion mentioned e.g. in Section \ref{Introduction}). This paper focuses not upon the steady state unsaturated line, but on the transient nutations, and their decay, described with $t_{DC},$ is produced by $s_zS_{jz}$ terms, and is therefore not linked with dipolar spectral diffusion. The situation found e.g. with $Mn^{2+}$ in ZnS is far more favourable for spectral diffusion, since in that case hyperfine coupling and spectral diffusion involve an electron and a nuclear spin on the same atom, and even then, transient experiments with fast field sweeps at $9$ $GHz,$ gave a spin diffusion time $T_{D}$ $\simeq$ $0.5$ $ms,$ at 4.2 and 1.34 K, while $T_1$ $=$ $0.27$ $s$ at 4.2 K and $0.88$ $s$ at 1.34 K \cite{Deville1975}. The decay of the TN is not linked {}with SL relaxation spectral diffusion either, as $T_{1}$ is presently assumed infinite. Under our assumptions, observation of spin diffusion in a transient experiment would therefore necessitate a time scale far longer than $t_{DC}$.  \\
In Eq. (\ref{RhoInitialGlobal}), $\mathcal{H}_{den}^{'}$ was neglected compared with  $\mathcal{H}_{Z}.$ If $\mathcal{H}_{den}^{'}$ is introduced, then   $\mathcal{H}_{Z}+\mathcal{H}_{den}^{'}$ $=\mathcal{H}_{Ze} + \mathcal{H}_{Zn}+\mathcal{H}_{den}^{'}$ is a sum of three Hamiltonians, each commuting with the two other ones. The existence of a common temperature $T$ is therefore not a trivial fact \cite{VanVleck1957}.\ The thermalization may result from a direct coupling of each term to a common bath, or from the neglected unsecular dipolar terms, $C,$ $E,$ and the $B$ term from $\mathcal{H}_{den}$.\ These unsecular terms, discussed in the context of NMR in \cite{Goldman1970}, are unable to play any role in the transient situation discussed in this paper.
\section{Conclusion}
Forced nutations of independent resonant A electron spins, and their decay induced by a weak dipolar coupling to non-resonant neighbouring B nuclear spins, were investigated. A general expression of the coherence time $t_{DC}$ was established. At exact ESR resonance, the dipolar coupling was found not to decohere the nutations. This led to seek the expression for $t_{DC}$ when $\Delta ,$ the shift from the resonance is equal to $\delta $, the half-width of the absorption line in steady state unsaturated ESR. The expression for $\delta$ was derived with the usual Van Vleck method of moments for undiluted solid samples, and its Kittel-Abraham extension for dilute samples. Numerical estimates for  $(t_{DC})_{\Delta =\delta }$ could be established for nuclear spins distributed over the sites of a simple cubic lattice. The qualitative results that $t_{DC}$ increases with the amplitude of the pulse inducing the nutations and shortens with $\Delta $ are reminiscent of those found in liquid state NMR in an inhomogeneous static field. Numerical estimations, with a possible interest both in the field of ESR and in the QIP context with electron spins as qubits,  were made.\\

\appendix{\hspace{-7mm} \Large \textbf{Appendix}}
\vspace{3mm}
\appendix{\large \textbf{A) Initial state and dipolar coupling}}\label{InitialStateAndDipCoupling} \\

One starts  \cite{Goldman1970} with Eq. (\ref{rhoZerolesdeuxàT}):
\begin{equation}
\rho(0)=\frac{\exp{[-(\mathcal{H}_Z+\mathcal{H}_{den}')/kT ]}}{\mathrm{Tr}\{\exp{[-(\mathcal{H}_Z+\mathcal{H}_{den}^{'}})/kT ]\}},
\end{equation}
where $\mathcal{H_Z}=\mathcal{H}_{Ze}+\mathcal{H}_{Zn}$, and $\mathcal{H}_{den}^{'}=\Sigma_{j}K_{0en}k_{j}I_{jz}s_{z}$.  The structure of the calculation leading to the first-order approximation for $\rho_{Pe}'(t)$ is unchanged. Keeping the HTA, the first-order expression for $\mathrm{Tr}\{\exp[-(H_{Z}+H_{den}^{'})/kT] \}$
is $2^{N+1}$, as $\mathrm{Tr}\{H_{den}^{'}\}=0$. One has therefore to replace Eq. (\ref{RhoInitialGlobal}) with:
\begin{equation}\label{Rho(0)Corrige} 
\rho_0\simeq \frac{1}{2^{N+1}}(1-\frac{\mathcal{H}_Z+\mathcal{H}_{den}^{'}}{kT}).
\end{equation}
One may guess that the contribution from $\mathcal{H}_{den}^{'}$ should be negligible compared with the one from $\mathcal{H}_Z$, because the weak dipolar coupling condition was assumed in Section \ref{SubsectionReducedDensityOp}. A more explicit discussion of that question is made here. Eq. (\ref{PartialTraceHauteTemp}) is then replaced with
\begin{equation}
\frac{1}{2^{N+1}}\mathrm{Tr}_{\mathcal{E}_n}\{e^{-i\tau_{eff}\widehat{\Sigma_z}}U_2
(1-\frac{\mathcal{H}_Z}{kT}-\frac{\mathcal{H}_{den}^{'}}{kT})U_2^\dagger e^{i\tau_{eff}\widehat{\Sigma_z}}\},
\end{equation}
where the $e^{-i\tau_{eff}\widehat{\Sigma_z}}$ operator originates from the dipolar term
$\mathcal{H}_{den}^{'}$. The first order contribution to $\rho_{Pe}^{'}(t)$ (cf. Eq. (\ref{ContientNewTrace}) from the dipolar term $\mathcal{H}_{den}^{'}$ in $\rho(0)$ is:
\begin{equation}
-\frac{1}{2^{N+1}kT}e^{\frac{-ih_{Z_{e,eff}}t}{\hbar}}\mathrm{Tr}_{\mathcal{E}_{n}} \{e^{-i\tau_{eff}\widehat{\Sigma_z}}U_2\mathcal{H}_{den}^{'} U_2^\dagger e^{i\tau_{eff}\widehat{\Sigma_z}}\}e^{\frac{ih_{Z_{e,eff}}t}{\hbar}}.
\end{equation}
This contribution is negligible compared with that from $\mathcal{H}_{Ze}/kT$, 
because $K_{0en}<<$ $\hbar \omega_1<<$ $\hbar \omega_0$. More explicitely, the contribution from 
$\mathcal{H}_{den}^{'}$ to the trace is a sum of contributions from each B spin. That from spin $N^{\circ} 1$ is:
\begin{equation}
K_{0en}k_1\mathrm{Tr}_{\mathcal{E}_n}\{e^{-iτ_{eff}\widehat{\Sigma_z}}U_2I_{1z}s_zU_2^{\dagger}e^{iτ_{eff}\widehat{\Sigma_z}}\}.
\end{equation}
$U_2$ transforms $I_{1z}s_z$ into a sum of four terms. The first one leads to a $I_{1z}s_z$ operator, 
unchanged by the $e^{-iτ_{eff}\widehat{\Sigma_z}}$ transformation, and finally giving no contribution to the 
first order correction, because $\mathrm{Tr}_{\mathcal{E}_{n}}I_{1z}$ $=0$. The second contribution from the spin B $N^{\circ} 1$ to the trace is with:	
\begin{equation}
-K_{0en}k_1\sin\Theta \cos \Theta \mathrm{Tr}_{\mathcal{E}_n}\{e^{-i\tau_{eff}\widehat{\Sigma_z}}s_xI_{1z}{e^{i\tau_{eff}\widehat{\Sigma_z}}}\}.
\end{equation}	
Using the kets $\mid m>$ $=\mid m_1>$..$\mid m_j>$..$\mid m_N>$, one may start the calculation with the determination of the partial trace within $\mathcal{E}_1$, the space state for spin B $N^{\circ}1$, which gives:
\begin{equation}
-K_{0en}k_1\sin\Theta\cos \Theta\sin(\frac{\tau_{eff}k_1}{2})\mathrm{Tr}_{\mathcal{E}_{2.N}}\{e^{-i \alpha}s_ye^{i \alpha}\}.
\end{equation}
In this trace within $\mathcal{E}_{2.N}$, the $I_{1z}$ operator is now absent, and the tracing, successively within $\mathcal{E}_2$, $\mathcal{E}_3$, down to  $\mathcal{E}_N$ , finally leads to the following contribution to $\rho_{Pe}^{'}(t)$:
\begin{equation}
\frac{K_{0en}k_1}{4kT}\sin\Theta\cos \Theta\sin\frac{\tau_{eff}k_1}{2}\prod_{j=2}^{n}\cos \frac{\tau_{eff}k_j}{2}s_y.
\end{equation}
We now know the contribution from this second term, and we observe that 1) this contribution cancels when
$\sin(\tau_{eff}k_1/2)=0$,  2) at the instants when both $\sin(\tau_{eff}k_1/2)$ and
$\prod_{j=2}^{n}\cos {\tau_{eff}k_j/2}$ are not negligible compared with 1, this contribution can be neglected compared with that from the terms which were kept in Section \ref{SubsectionReducedDensityOp}, because the weak dipolar coupling condition and the HTA both have been assumed: $K_{0en}<<\hbar \omega_1<<\hbar \omega_0<<kT$.\\
The third and fourth terms involve a $I_{1x}$ operator. Keeping the same approach, one first calculates the partial trace within  $\mathcal{E}_1$. But $<\pm1/2\mid I_{1x}\mid\pm1/2>=0$, and therefore the third and fourth terms have no contribution to $\rho_{Pe}^{'}(t)$.\\

\appendix{\large \textbf{B) On the path to the reduced density operator}}\label{OnThePathToReducedDensOp} \\

It is presently supposed that $\Sigma_e$ and $\Sigma_n$ are initially both at thermal equilibrium. The contribution from the $-\sin \Theta s_{x}$ term to Eq. (\ref{ZeemanSpinAExpress1}) is:
\begin{equation}\label{IntroductiondeT} 
\frac{\hbar \omega_0}{2^{N+1}kT}\sin \Theta \mathrm{Tr}_{\mathcal{E}_n}\{e^{-i \tau_{eff}\widehat{\Sigma_z}}s_{x}e^{i\tau_{eff}\widehat{\Sigma_z}}\},
\end{equation}
with $\widehat{\Sigma_z}=$ $\Sigma_{j}k_{j}I_{jz}s_{z}$. Choosing to calculate the trace within the $2^N$ kets $\mid m>$ $=$\medskip $\mid m_1>$..$\mid m_j>$..$\mid m_N>$ of the standard basis, we first consider the following diagonal element, where $m_1=1/2$:
\begin{eqnarray}
D_{1+}&=&<m_N,.m_j,.\frac{1}{2}\mid\hat{ O_1}\mid \frac{1}{2},.m_j,.m_N>,\\
&=&<m_N,.m_j,.m_2 \mid \hat{ A}_2\mid  m_2,.m_j,.m_N>,
\end{eqnarray}
with
\begin{eqnarray}
\hat{ O_1}&=&e^{-i \tau_{eff}\widehat{\Sigma_z}}s_xe^{i \tau_{eff} \widehat{\Sigma_z}}, \\
\hat{ A}_2&=&e^{-i \alpha}(\cos\frac{\tau_{eff}k_1}{2} s_x +\sin\frac{\tau_{eff}k_1}{2} s_y) e^{i \alpha}, \\
\alpha &=&\tau_{eff} \Sigma _{j=2}^{N}k_jI_{jz}s_z.
\end{eqnarray}
With the diagonal element $D_+$, we associate the following diagonal element ($m_1=-1/2$):
\begin{equation}
D_-=<m_N,.m_j,.-\frac{1}{2}\mid \hat{ O_1} \mid -\frac{1}{2},.m_j,.m_N>.
\end{equation}
The partial trace $\mathrm{Tr}_{\mathcal{E}_1}$ within the space state of this first B type spin, equal to 
$D_+$ + $D_-$, does not contain the $s_y$ part and leaves us with:
\begin{equation}
2 \cos(\frac{\tau_{eff}k_1}{2})\mathrm{Tr}_{\mathcal{E}_{2.N}}\{e^{-i \alpha}s_ze^{i \alpha}\},
\end{equation}
where the trace $\mathrm{Tr}_{\mathcal{E}_{2.N}}$ is restricted to the state space of the B spins numbered from 2 to N. It is then possible to repeat the same process within the state space of spin $N^{\circ} 2$, and this again down to spin N, finally leading to the following expression for the trace in Eq. (\ref{IntroductiondeT}):
\begin{equation}
2^{N} \Pi_{j=1}^{N} \cos \frac{\tau_{eff}k_j}{2},
\end{equation}
and Eq. (\ref{IntroductiondeT}) is therefore equal to:
\begin{equation}
\frac{\hbar \omega_0}{2kT}\sin\Theta (\Pi_{j=1}^{N}\cos\frac{\tau_{eff}k_j}{2})s_x.
\linebreak 
\end{equation}

\appendix{\large \textbf{C) Moments of the resonance line}}\label{MomentsResonanceLine} \\

In Section \ref{TransientBehavior}, a model of independent clusters is used, and therefore when calculating moments of the resonance line, one may consider a single cluster. The second moment $M_2^{'}$ of the line \cite{Abragam1961} is proportional to 
$\mathrm{Tr}\{[\mathcal{H}_{den}^{'},s_x]\}$ (cf. Eq. (\ref{SecondMoment})). From Eqs. (\ref{HamiltonienDeDépart}), (\ref{DipolairesA}), (\ref{HdPrime1}) and (\ref{K0eS}), the commutator is
\begin{equation}\label{commutDipsSAvecsx}
[\mathcal{H}_{den}^{'},s_x]=iK_{0en}s_y \Sigma_jk_jI_{jz} , 
\end{equation}
and therefore:
\begin{equation}
\hbar^{2}M_2^{'}=\frac{K_{0en}^{2}\mathrm{Tr}_{\mathcal{E}_j}\{I_{jz}^{2}\}\Sigma_jk_j^{2}}{(2I+1)},
\end{equation}
(${\mathcal{E}_j}$: state space for spin B $N^{\circ}$ j), which leads to Eq. (\ref{SecondMomentf=1}).\\
From arguments given in Section \ref{TransientBehavior}, the fourth moment $M_4'$ reduces to:
\begin{equation}\label{MPrime4Appendix} 
\hbar^{4}M_4^{'}=
\frac{\mathrm{Tr}\{[\mathcal{H}_{den}^{'},[\mathcal{H}_{den}^{'},s_x]]^{2}\}}{\mathrm{Tr}\{s_x^{2}\}},
\end{equation}
as the secular part of the dipolar coupling is presently equal to $\mathcal{H}_{den}^{'}$. The commutator $[\mathcal{H}_{den}^{'},[\mathcal{H}_{den}^{'},s_x]]$ equals:
\begin{equation}
K_{0en}^{2}[s_z\sum_jk_jI_{jz},is_y\sum_lk_lI_{lz}]=K_{0en}^{2}s_x\sum_{jl}k_jk_lI_{jz}I_{lz},
\end{equation}
with j and l different or not, and its square equals:
\begin{equation}\label{CarreCommuTMoment4Append} 
K_{0en}^{4}s_x^{2}\sum_{jlmn}k_{j}k_{l}k_{m}k_{n}I_{jz}I_{lz}I_{mz}I_{nz},
\end{equation}
with j,l, m, and n taking all the integer values from 1 to N. When tracing this sum, all the terms with three distinct indices or with three equal indices cancel, as $\mathrm{Tr}\{I_{jz}\}=0$. For each  j value there is a single term $k_j^{4}I_{jz}^{4}$, and the total contribution from these terms to the numerator of 
Eq. (\ref{MPrime4Appendix}) is:
\begin{equation}\label{Moment4K4} 
K_{0en}^{4}(\sum_{j}k_{j}^{4})\mathrm{Tr}\{s_{x}^{2}I_{jz}^{4}\}.
\end{equation}
In Eq. (\ref{CarreCommuTMoment4Append}), there are three sums arising respectively from $k_j^{2}k_l^{2}$ terms, from $k_j^{2}k_m^{2}$ terms and from $k_j^{2}k_n^{2}$ terms. Each sum gives the same contribution to the numerator of Eq. (\ref{MPrime4Appendix}). Their total contribution is:
\begin{equation}\label{Moment4LJ2KL2} 
3K_{0en}^{4}(\sum_{j,l\ne j}k_{j}^{2}k_{l}^{2})\mathrm{Tr}\{s_{x}^{2}I_{jz}^{2}I_{lz}^{2}\}.
\end{equation}
Inserting (\ref{Moment4K4}) and (\ref{Moment4LJ2KL2}) in (\ref{MPrime4Appendix}) gives:
\begin{equation}
\hbar^{4}M_4^{'}=\frac{K_{0en}^{4}}{(2I+1)^{2}}[(\sum_{j}k_j^{4})\mathrm{Tr}_{\mathcal{E}_j}\{I_{jz}^{4}\}(2I+1)+3(\sum_{j,l\ne j}k_j^{2}k_l^{2})(\mathrm{Tr}_{\mathcal{E}_j}\{I_{jz}^{2}\})^{2}].
\end{equation}
Calculating the traces and defining $f(I)=$ $(3I^2+3I-1)$ and $g(I)=$ $I(I+1)$, one thus gets for $M_4^{'}$:
\begin{equation}\label{MPrime4General} 
\hbar^{4}M_4^{'}=\frac{K_{0en}^{4}I(I+1)}{15}[\sum_{j}k_j^{4}f(I)+5\sum_{j,l\ne j}k_j^{2}k_l^{2}g(I)]
\end{equation}
Presently, I=1/2, and Eq. (\ref{MPrime4General}) leads to Eq. (\ref{MPrime4fegal1}). The sums $\sum_{j}k_{j}^{4}$ and $\sum_{jl}k_{j}^{2}k_{l}^{2}$ have then to be explicited in the presently chosen case (simple cubic lattice, static field along a four-fold symmetry axis). Their numerical values are $\sum_{j}k_{j}^{4}$ $\simeq≃36.1$ and $\sum_{j,l\neq j}k_{j}^{2}k_{l}^{2}$ $\simeq142.1$.

\end{document}